%% file: main.tex
\documentclass[acmsmall,screen]{acmart}

\usepackage{pifont}% http://ctan.org/pkg/pifont
\usepackage{color,xcolor}
\usepackage{multirow}
\usepackage{listings}
\usepackage[shortlabels]{enumitem}
\usepackage[most]{tcolorbox} 
\usepackage{array}
\usepackage{tabularx}
\usepackage{graphicx}
\usepackage{hyperref}
\usepackage{float}
\usepackage{algpseudocode}
\usepackage{soul}
\usepackage{wrapfig}
\usepackage{subcaption}
\usepackage{tikz}

\usetikzlibrary{positioning, shapes.geometric, arrows.meta}

\AtBeginDocument{%
  }

\newcommand{\parh}[1]{\noindent\textbf{#1}}

\newcommand{\sm}{Supplementary Material}

\newcommand{\F}{Fig.}

\newcommand{\T}{Table}

\newcommand{\finding}[2]{
\begin{tcolorbox}[width=\linewidth,boxrule=1pt,top=1pt, bottom=1pt, left=1pt,right=1pt, colback=white!20,colframe=black!20]
\textbf{Finding #1:} %\textit
{#2}
\end{tcolorbox}}
\newcommand{\answer}[2]{
\begin{tcolorbox}[width=\linewidth,boxrule=1pt,top=1pt, bottom=1pt, left=1pt,right=1pt, colback=white!20,colframe=black!20]
\textbf{Answer to RQ#1:} %\textit
{#2}
\end{tcolorbox}}
\newcommand{\autour}[1]{\tikz[baseline=(X.base)]\node [draw=none,fill=gray!15,semithick,rectangle,inner sep=2pt, rounded corners=3pt] (X) {\texttt{#1}};}

\begin{document}

\title{From Evaluation to Enhancement: Large Language Models for Zero-Knowledge Proof Code Generation}

\author{Zhantong Xue}
\orcid{0009-0000-3419-9431}
\affiliation{%
  \institution{Hong Kong University of Science and Technology}
  \city{Hong Kong}
  \country{China}
}
\email{zxueai@cse.ust.hk}

\author{Pingchuan Ma}
\authornote{Corresponding author}
\orcid{0000-0001-7680-2817}
\affiliation{%
  \institution{Zhejiang University of Technology}
  \city{Hangzhou}
  \country{China}
}
\affiliation{%
  \institution{CipherInsight Limited}
  \city{Hong Kong}
  \country{China}
}
\email{pma@zjut.edu.cn}

\author{Zhaoyu Wang}
\orcid{0009-0009-6892-1264}
\affiliation{%
  \institution{Hong Kong University of Science and Technology}
  \city{Hong Kong}
  \country{China}
}
\email{zwangjz@cse.ust.hk}

\author{Yuguang Zhou}
\orcid{0000-0001-5069-4563}
\affiliation{%
  \institution{Hong Kong University of Science and Technology}
  \city{Hong Kong}
  \country{China}
}
\email{yzhougv@cse.ust.hk}

\author{Xiaoqin Zhang}
\orcid{0000-0003-0958-7285}
\affiliation{%
  \institution{Zhejiang University of Technology}
  \city{Hangzhou}
  \country{China}
}
\email{xqzhang@zjut.edu.cn}

\author{Shuai Wang}
\orcid{0000-0002-0866-0308}
\affiliation{%
  \institution{Hong Kong University of Science and Technology}
  \city{Hong Kong}
  \country{China}
}
\affiliation{%
  \institution{CipherInsight Limited}
  \city{Hong Kong}
  \country{China}
}
\email{shuaiw@cse.ust.hk}

\author{Juergen Rahmel}
\orcid{0009-0007-5080-8736}
\affiliation{%
  \institution{HSBC}
  \city{Hong Kong}
  \country{China}
}
\email{juergen.rahmel@hsbc.com.hk}

% remove the copyright information
\setcopyright{none}
\settopmatter{printacmref=false} % Removes citation information below abstract
\renewcommand\footnotetextcopyrightpermission[1]{} % removes footnote with conference information in first column

% \renewcommand{\shortauthors}{Authors et al.}


\begin{abstract}
Zero-knowledge proofs (ZKPs) are increasingly deployed in domains such as
privacy-preserving authentication, verifiable computation, and secure finance.
However, authoring ZK programs remains challenging: unlike conventional software
development, ZK programming manifests a fundamental paradigm shift from
\textit{imperative computation} to \textit{declarative verification}. This
process requires rigorous reasoning about finite field arithmetic and complex
constraint systems (which is rare in common imperative languages), making it
knowledge-intensive and error-prone. While large language models (LLMs) have
demonstrated strong code generation capabilities in general-purpose languages,
their effectiveness for ZK programming, where correctness hinges on both
language mastery and constraint-level reasoning, remains unexplored. To address
this gap, we propose \textsc{ZK-Eval}, a domain-specific evaluation pipeline
that probes LLM capabilities on ZK programming at three levels: language
knowledge, algebraic primitive competence, and end-to-end program generation.
Our evaluation of four state-of-the-art LLMs reveals that while models
demonstrate strong proficiency in language syntax, they struggle when
implementing and composing algebraic primitives to specify correct constraint
systems, frequently producing incorrect programs. Based on these insights, we
introduce \textsc{ZK-Coder}, an agentic framework that augments LLMs with
constraint sketching, guided retrieval, and interactive repair. Experiments with
GPT-o3 on Circom and Noir show substantial gains, with success rates improving
from 20.29\% to 87.85\% and from 28.38\% to 97.79\%, respectively. With
\textsc{ZK-Eval} and \textsc{ZK-Coder}, we establish a new basis for
systematically measuring and augmenting LLMs in ZK code generation to lower
barriers for practitioners and advance privacy computing.
\end{abstract}



%
% Keywords. The author(s) should pick words that accurately describe
% the work being presented. Separate the keywords with commas.
% \keywords{Zero-Knowledge Proofs, Large Language Models, Code Generation}

\received{30 January 2026}

\maketitle

\input{introduction}
\input{background}
\input{capability-study}
\input{augmenting}
\input{evaluation}
\input{related}
\input{discussion}
\input{conclusion}

\bibliographystyle{ACM-Reference-Format}
\bibliography{main}

% \clearpage
% \appendix

\input{illustrative-workflow}
\input{real-world-coverage}
\input{case-studies-public-repo}
\input{filtered-tasks}
\input{prompts-available}

\end{document}

%% file: introduction.tex
\section{Introduction}
\label{sec:introduction}

\emph{Zero-knowledge proofs (ZKPs)} are a powerful cryptographic primitive that
allow one party to prove knowledge of a statement without revealing the
underlying witness. In practice, this means proving that a computation or claim
is correct while keeping sensitive inputs private. ZKPs have moved far beyond
theory: they are now central to privacy-preserving authentication, blockchain
scalability, verifiable computation, and emerging applications across finance
and security domains~\cite{sun2024zkllmzeroknowledgeproofs, Gupta2025ZKPSurvey,
lavin2024survey, Tangem2024ZKPBeyondPrivacy}. As demand for trustworthy
computation grows, the ability to author, test, and verify ZKP programs has
become a first-class concern in the software engineering
community~\cite{takahashi2025zkfuzz, hochrainer2024fuzzing, xu2025towards,
pailoor2023automated, jiang2025conscs}.

\begin{figure}[ht]
    \centering
    \includegraphics[width=\linewidth]{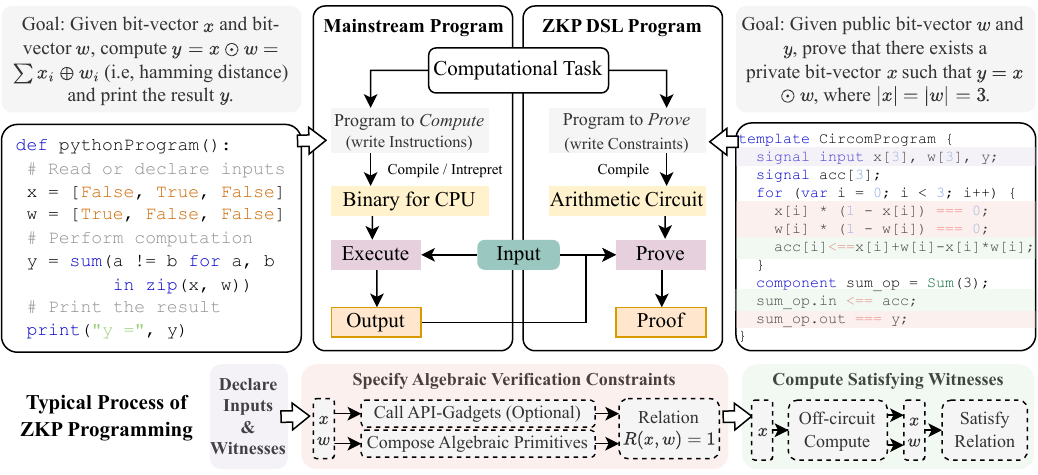}
    \caption{Illustrative Comparison Between the Process of Mainstream and ZK Programming.}
    \label{fig:illustrative-programming-process-comparison}
    \vspace{-5pt}
\end{figure}

However, authoring ZK programs is challenging. Unlike mainstream programming,
which builds on the von Neumann model and imperative execution, ZKP programs
boil down to specifying and proving mathematical \emph{constraints} over finite
fields. In essence, this represents a paradigm shift from computation to
verification. Its development requires reasoning about finite field arithmetic
and constraint systems. As shown in
\F~\ref{fig:illustrative-programming-process-comparison} and
\T~\ref{tab:zkp-conceptual-differences}, ZK programs do not describe how to
\emph{execute} a computation but how to \emph{prove} its correctness by
compiling constraints into arithmetic circuits. Developers must specify
equations over inputs, outputs, and intermediate variables, carefully wiring
algebraic primitives to specify constraints enforcing correctness. Even small
omissions or miswirings can yield unsound or incomplete constraint systems,
leading to errors that compilers rarely flag. This makes ZK development
knowledge-intensive, error-prone, and inaccessible to non-experts, despite
significant progress in proof systems and verification
tools~\cite{chaliasos2024sok, pailoor2023automated, takahashi2025zkfuzz,
kolozyan2025language, chaliasos2025towards}.

\begin{figure}[t]
    \centering
    \begin{minipage}{.47\textwidth}
        \input{mainstream-zk-comparison}
    \end{minipage}
    \hspace{0.02\textwidth}
    \begin{minipage}{.48\textwidth}
        \centering
        \includegraphics[width=\linewidth]{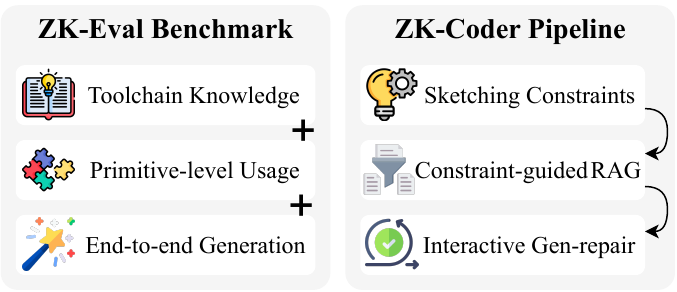}
        \caption{Overview of \textsc{ZK-Eval} and \textsc{ZK-Coder}.}
        \label{fig:high-level-pipeline}
    \end{minipage}
    \vspace{-8pt}
\end{figure}

As a result, ZK programming demands significantly more time and expertise than
mainstream software development, posing a steep barrier to entry for many
practitioners. In contrast, recent advances in large language models (LLMs) have
lowered this barrier in general-purpose languages by demonstrating strong code
generation capabilities~\cite{10.1145/3747588}, achieving impressive results on
benchmarks such as HumanEval~\cite{chen2021evaluating} and
MBPP~\cite{mbppdataset}, and powering widely adopted tools like GitHub
Copilot~\cite{copilot} and Cursor~\cite{cursor}. This contrast naturally raises
the question: \emph{to what extent can modern LLMs assist in generating ZK code
from natural language specifications?} Answering this question requires more
than surface-level benchmarking. Compared to programming in mainstream
languages, success in ZK development hinges on three competencies: (i) knowledge
of domain-specific languages (DSLs) and their toolchains, (ii) mastery of
algebraic primitives as atomic building blocks for constraint systems, and (iii)
proficiency in modeling and specifying complex verification constraints.
Existing benchmarks focus only on mainstream programming paradigms and overlook
these requirements.

To measure this knowledge gap, this paper first introduces \textsc{ZK-Eval}, a
domain-specific evaluation pipeline that probes LLM capability at three levels:
language knowledge, primitive-level competence, and end-to-end program code
generation (left panel of \F~\ref{fig:high-level-pipeline}). We anticipate that
this design would provide a structured way to pinpoint where models succeed,
where they fail, and why.

As will be detailed in \S~\ref{sec:capability-study}, applying \textsc{ZK-Eval}
to four SoTA LLMs (including advanced reasoning models like GPT-o4-mini and
GPT-o3) reveals a clear pattern. While models perform strongly on language
knowledge, their accuracy drops significantly (below 30\%) when reasoning about
primitives or assembling verification constraints, highlighting a critical gap
between syntactic knowledge and end-to-end generation. These results underscore
both the \emph{potential} of LLMs to democratize ZK development and the
\emph{need} for targeted augmentations to bridge this capability gap. Motivated
by these findings, we present \textsc{ZK-Coder} (right panel of
\F~\ref{fig:high-level-pipeline}), which combines three key components: (i) a
sketch layer (ZKSL) that explicitly models constraints using algebraic
primitives, (ii) retrieval-augmented generation to ground specifications in
verified implementations, and (iii) interactive refinement to fix constraint
misspecifications. We observe that \textsc{ZK-Coder} achieves substantial gains,
improving code generation success rates from 20.29\% to 87.85\% on Circom and
from 28.38\% to 97.79\% on Noir (using GPT-o3). Ablation studies further confirm
that sketching, guided retrieval and interactive refinement are indispensable;
removing either component causes a significant performance drop (over 10\%). In
summary, our contributions are as follows:
\begin{enumerate}[leftmargin=*]
    \item We propose \textsc{ZK-Eval}, a three-stage evaluation pipeline for
    systematically assessing LLM capabilities in ZK code generation.
    \item We introduce \textsc{ZK-Coder}, an agentic framework that augments
    LLMs with constraint sketching, guided retrieval, and interactive repair to
    improve reliability.
    \item We evaluate the performance of four state-of-the-art LLMs on
    \textsc{ZK-Eval} and demonstrate how \textsc{ZK-Coder} substantially
    improves end-to-end code generation.
\end{enumerate}

%% file: mainstream-zk-comparison.tex
\captionof{table}{Mainstream \& ZK Program Comparison.}
\label{tab:zkp-conceptual-differences}
\resizebox{\linewidth}{!}{
    \begin{tabular}{ccc}
    \toprule
    \textbf{Aspect} & \textbf{Mainstream} & \textbf{ZK Program} \\
    \midrule
    \textbf{Paradigm} & Imperative/OOP/Func & Constraint Systems \\
    \textbf{Comp. Model} & von Neumann Arch & Arithmetic Circuits \\
    \textbf{Purpose} & to Compute & to Prove/Verify \\
    \textbf{Data Types} & Int/Float/Str/... & Finite Field Elements  \\
    \textbf{Mutability} & Dynamic & Static \\
    \textbf{Control Flow} & If/Loop/Recursion & Limited \\
    \textbf{Side Effects} & Yes & No, Pure Math \\
    \bottomrule
    \end{tabular}
}

%% file: background.tex
\section{Background}
\label{sec:background}

\subsection{Zero-Knowledge Proofs and Programming Frameworks}
\label{subsec:zkp-frameworks}

Zero-knowledge proofs (ZKPs) are cryptographic protocols that allow one party to
prove to another that a statement is true without revealing any information
beyond the validity of the statement itself. Formally, the following properties
are desired for a correct ZKP program.
\begin{enumerate}[leftmargin=*]
    \item \textbf{Completeness:} If a statement is true, an honest prover can
    convince the verifier of its validity.
    \item \textbf{Soundness:} If a statement is false, no dishonest prover can
    convince the verifier of its validity.
    \item \textbf{Zero-Knowledge:} The verifier learns nothing beyond the
    validity of the statement itself.
\end{enumerate}
These properties make ZKPs particularly useful in verifiable computation and
privacy-preserving applications~\cite{Gupta2025ZKPSurvey,
Tangem2024ZKPBeyondPrivacy,lavin2024survey}. However, ZK programming
necessitates a fundamental paradigm shift from imperative computation to
declarative verification. Rather than writing instructions to calculate a
result, developers must model their high-level intent as an arithmetic
representation $R(x, w) = 1$ over a finite field, where $x$ denotes public
inputs and $w$ denotes the private witness. This requirement to faithfully
enforce logical intent through field-arithmetic constraints is notoriously
arduous for humans; even minor errors in modeling can lead to under-constrained
circuits (compromising soundness) or over-constrained systems (breaking
completeness)~\cite{pailoor2023automated, wen2024practical, chaliasos2024sok,
kolozyan2025language}.

To mitigate this complexity, several programming frameworks have emerged,
offering a diverse range of abstractions. We categorize these into two primary
design paradigms:

\begin{enumerate}[leftmargin=*]
    \item \textbf{Constraint-Oriented DSLs.} Frameworks like Circom and Noir
    provide specialized syntax for expressing verification relations. Circom
    requires developers to think in terms of low-level circuit wiring and
    explicit signal management, which poses significant challenges for LLMs
    tasked with producing correct circuit templates~\cite{hochrainer2024fuzzing,
    chaliasos2025towards}. Conversely, Noir adopts a Rust-like syntax with
    strong typing, aligning more closely with the general-purpose coding corpora
    on which current LLMs are trained~\cite{joel2024survey, jiang2024survey}.
    \item \textbf{General-Purpose zkVMs.} Emerging solutions like SP1~\cite{sp1}
    and RISC0~\cite{risczero} allow developers to compile standard Rust code
    into a restricted instruction set, generating proofs over the execution
    trace within a virtual machine. While they abstract away the constraint
    system entirely, they often incur substantial performance overhead. We
    consider these an orthogonal approach to our work, as we focus on the direct
    generation of circuits.
\end{enumerate}

In this work, we focus on Circom and Noir because they represent the most widely
adopted frontends in both academia~\cite{stronati2024clap, noirlang,
ozdemir2022circ, pailoor2023automated, wen2024practical} and industry (e.g., in
Ethereum rollups and Aztec). Furthermore, they occupy two complementary points
on the abstraction spectrum: Circom tests an LLM's ability to perform
circuit-level specification and constraint wiring, while Noir evaluates how
effectively models can adapt conventional coding proficiency to the
verification-oriented paradigm. By studying both, we capture the core technical
challenges of ZKP development while ensuring our findings have immediate
practical relevance.

\subsection{LLM-based Code Generation}

Large Language Models (LLMs) have emerged as powerful assistants in software
development, enabling code generation from natural language prompts with tools
like GitHub Copilot, Cursor, and model-driven agents quickly translating intent
into code~\cite{jiang2024survey,10.1145/3747588}. These systems are reshaping
engineering workflows, aiding in writing, testing, and documenting code by
drawing on vast amounts of publicly available repositories and
documentation~\cite{copilot}.

\noindent \textbf{Framework.}~Modern LLM-based code generation typically follows
an agentic framework~\cite{zhang2024codeagent, wang2025agents}, where the model
acts as an intelligent assistant, interpreting user intent, retrieving relevant
information, and producing code snippets or entire functions as needed. This
process often involves multiple iterations, with the model refining its output
based on user feedback or additional context. In this way, it can leverage
off-the-shelf LLMs effectively without code-specific
fine-tuning~\cite{dong2025survey, pan2025codecor}. However, to the best of our
knowledge, there has been limited exploration of LLMs for code generation for
low-resource programming languages (LRPL) or DSLs especially for ZKP DSLs.

\noindent \textbf{Benchmark.}~Benchmarks are a crucial component of LLM-based
code generation research, playing a vital role in understanding model
performance and guiding improvements. Currently, their accuracy and reasoning
ability are usually evaluated on standard benchmarks on mainstream programming
languages such as HumanEval~\cite{chen2021evaluating} and
MBPP~\cite{mbppdataset}. However, evaluation tools for LRPLs and DSLs remain
limited, hindering comprehensive assessments of LLM performance in these
contexts~\cite{joel2024survey, giagnorio2025enhancing}. Code generation for ZKP
programs exemplifies the challenges and opportunities in this space. Though
benchmarks for LRPLs~\cite{cassano2022multiplescalableextensibleapproach,
yang2023intercode} and DSLs~\cite{liu2023verilogeval, abukhalaf2023codex} exist,
to our knowledge, no standardized benchmark suite specifically targets ZKP code
generation, requiring tailored benchmarks and evaluation strategies.

%% file: capability-study.tex
\section{Measuring LLM Capability for ZK Code Generation}
\label{sec:capability-study}

As discussed in \S~\ref{sec:introduction} and \S~\ref{subsec:zkp-frameworks}, ZK
programming follows a proof-oriented model: developers translate high-level
intent into algebraic constraints over a finite field and produce proofs of
satisfiability. This shift yields a unique programming workflow depicted in
\F~\ref{fig:illustrative-programming-process-comparison}, where developers
construct the verification relation by wiring together algebraic primitives
(mathematical gates that enforce logical invariants over finite fields). While
witness computation still resembles conventional imperative programming, the
developer's primary responsibility shifts to the rigorous specification of these
verification constraints. To evaluate LLM proficiency in this critical domain,
we assess two foundational capabilities essential for end-to-end success:
\ding{192} language and toolchain knowledge and \ding{193} algebraic primitive
competence.

\begin{enumerate}[leftmargin=*]
    \item \textbf{Language and Toolchain Knowledge.} Developers must master
    domain-specific syntax, strict typing rules, and the semantics of field
    operations to structure valid constraint specification and witness
    assignments. Weakness in any of these areas routinely leads to programs that
    cannot be compiled or produce an invalid relation that renders the proof
    meaningless.
    \item \textbf{Algebraic Primitive Competence.} Algebraic primitives are the
    atomic building blocks of \textit{ZK constraint systems}. Since arithmetic
    circuits lack native support for high-level operations, developers combine
    relational (e.g., comparators), logical (e.g., bitwise gates), and
    arithmetic (e.g., division, inversion) primitives to express logic.
    Competence in these atomic units is a prerequisite for constructing sound,
    higher-level application-specific constraint systems.
\end{enumerate}

These capabilities constitute two core dimensions of end-to-end ZK program
generation. Without language and toolchain knowledge, LLMs cannot produce
compilable programs; without algebraic primitive competence, programs cannot
faithfully encode intended relations. Since ZK programming requires integrating
language-level fluency with deeper constraint modeling, existing mainstream
benchmarks~\cite{chen2021evaluating, jain2024livecodebench, mbppdataset} leave
this integration unexplored. We introduce \textsc{ZK-Eval} to systematically
measure both dimensions and their integration in end-to-end tasks.

\subsection{ZK-Eval Benchmark}
\label{subsec:zk-eval-design}

\begin{figure*}[ht]
    \centering
    \includegraphics[width=\textwidth]{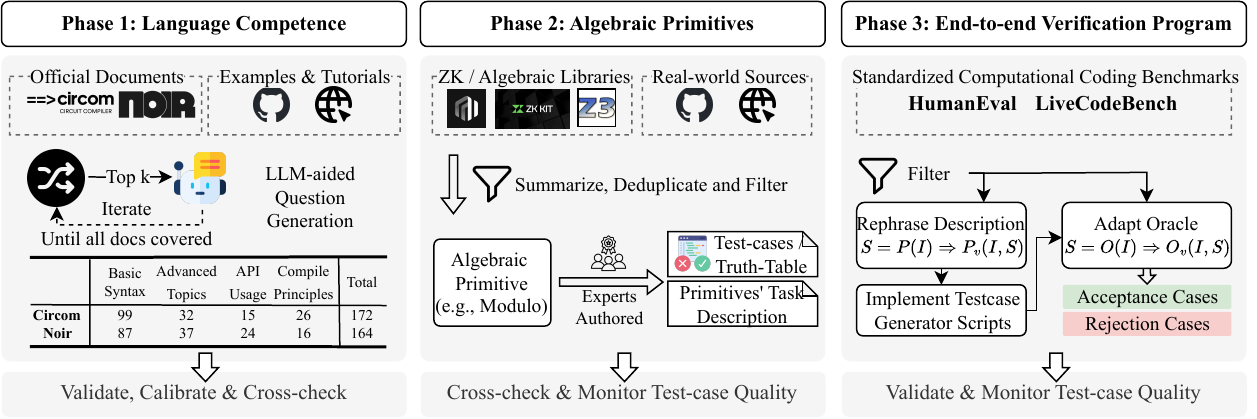}
    \caption{The Three-Stage Evaluation Design of the \textsc{ZK-Eval} Benchmark.}
    \label{fig:zk-eval-overview}
    \vspace{-10pt}
\end{figure*}

To realize this benchmark design, we structure \textsc{ZK-Eval} in three stages,
each targeting one capability and their integration. We illustrate the design
overview in \F~\ref{fig:zk-eval-overview}.

\subsubsection{Language and Toolchain Knowledge}
\label{subsubsec:zk-eval-knowledge}

We first evaluate whether LLMs possess the foundational knowledge required to
author ZK programs. As shown in the left side of \F~\ref{fig:zk-eval-overview},
sources are collected through web crawling of official documentation, including
language specifications, tutorials, and API references (e.g., Circom's
\texttt{circomlib}~\cite{circomlib} and the Noir's \texttt{stdlib}).
Additionally, we gather publicly available pedagogical examples from GitHub
repositories~\cite{circomkitexamples, circomexample, noirexamples,
circuitexamples} and online tutorials~\cite{circomerhant} to increase coverage
of idiomatic usage. This yields 51 documents for Circom and 40 for Noir.

Based on this knowledge base, we then design a pipeline to automatically generate
multiple-choice questions, a format widely adopted in prior evaluations of
language models~\cite{hendrycks2020measuring,myrzakhan2024open,
zellers2019hellaswag,clark2018think}. Inspired by earlier work on LLM-driven
question generation~\cite{liu2020asking,abbasiantaeb2024let,long2024llms,
ji2025measuring}, our pipeline iteratively samples subsets of documents and
provides them to a large model, which composes candidate questions and answers.
This process is repeated until the full document collection has been
exhaustively covered, producing a diverse set of drafts.

All generated questions are then subjected to rigorous expert validation. Three
human reviewers independently assess each item for factual correctness, clarity,
and alignment with the intended difficulty, with disagreements resolved through
discussion to consensus. Cross-checking by the authors ensures consistency
across the set. Topic coverage is also verified to ensure that all major aspects
(as listed in the official documents) of the studied language are represented.
We categorize the questions into four groups: \emph{basic syntax} (language
grammar such as datatypes, statements, operators, control flows, signals and
buses), \emph{advanced topics} (e.g., taggings, traits, generics), \emph{API
references} (standard library APIs), and \emph{compiler principles} (translation
from source code to arithmetic circuits). As shown in
\F~\ref{fig:zk-eval-overview}, the final benchmark consists of 172 questions for
Circom and 164 for Noir, with a balanced distribution across the four
categories. Representative examples are provided in the \sm~and the full set of
questions is provided in the Artifact.

\subsubsection{Implementing and Using Algebraic Primitives to Specify Constraints}
\label{subsubsec:zk-eval-primitives}

We assess the capability of LLMs to implement and utilize algebraic primitives
as the building blocks of ZK constraint systems. Different from higher-level
cryptographic (e.g., hash functions, signatures) and application-level (e.g.,
account-balance checks, order-matching validity) APIs available in external
libraries, these encapsulate the core relational logic that developers must wire
to define custom circuit behavior.

\noindent \textbf{Technical Difficulty.} We clarify that, constructing a
benchmark that reflects the fundamental algebraic patterns essential to
real-world ZK programs is \textit{challenging}. In existing codebases, core
relational logic is often entangled with application-specific boilerplate,
making it difficult to isolate and evaluate the underlying algebraic reasoning.
To decouple and isolate these foundational patterns, we systematically curate a
set of primitives from widely adopted libraries. This process, illustrated in
\F~\ref{fig:zk-eval-overview}, draws from three complementary sources:
\texttt{circomlib}~\cite{circomlib}, \texttt{zk-kit}~\cite{zkkitrepo}, and
Z3Py~\cite{z3py}.

\begin{enumerate}[leftmargin=*]
    \item \textbf{Standardized Libraries.} We first include canonical primitives
    from \texttt{circomlib} and \texttt{zk-kit}, which represent the production
    standard for routine circuit instantiation.
    \item \textbf{Agnostic Extraction.} From these repositories, we manually
    extract and deduplicate algebraic primitives while filtering out
    application-specific composites. This ensures the benchmark captures the
    universal building blocks of ZK logic rather than application-specific
    patterns.
    \item \textbf{Aligned Completeness.} To ensure comprehensiveness, we
    cross-reference our set with the \texttt{Z3Py} SMT-solver interface and
    incorporate primitives with similar structures, guaranteeing completeness
    for algebraic constraint specification.
\end{enumerate}

The resulting benchmark consists of 35 algebraic primitives covering a wide
range of categories: (i) \emph{basic checks}, including range tests and
comparisons; (ii) \emph{logical gates}, such as AND, OR, XOR, and equality;
(iii) \emph{arithmetic operators}, such as addition, multiplication, integer
division, exponentiation, modular reduction, and inversion; and (iv)
\emph{composite operators}, such as sum and multiplexers (see the full list in
the \sm). This collection serves as a complete basis for constraint
specification; as these algebraic primitives represent the atomic operations of
arithmetic circuits, any non-trivial verification relation can be realized
through their rigorous composition. 

\noindent \textbf{Practical Relevance.}~To check the practical relevance of our
benchmark, we surveyed 37 production-grade ZK repositories comprising 408 source
files. Our analysis confirms that the selected algebraic primitives are highly
representative of real-world circuit logic: we achieve 97.30\% and 94.59\%
coverage for core arithmetic and relational primitives, respectively, which are
universal to production ZK systems. While logical (70.27\%) and specialized
composites (16.22\%) appear less frequently, they remain mathematically
essential for logical completeness and the construction of complex verification
relations. Notably, it includes every one of the top 10 most frequently used
primitives identified in the survey, establishing a representative foundation.

\noindent \textbf{Task Setup.}~For each primitive, we author a concise task
description that asks the LLM to implement the primitive's functional logic and
use it to specify the intended algebraic relation. Following established
practice on test-case curation~\cite{chen2021evaluating, jain2024livecodebench},
we craft test-cases that validate both completeness (accepting valid witnesses)
and soundness (rejecting invalid assignments). For finite primitives (e.g.,
logical xor), test-cases exhaustively enumerate the entire truth table. For
primitives with larger domains, we combine automated random input generation
with edge-cases (e.g., field overflows, zero-value or out-of-domain inputs),
yielding at least 100 cases per primitive. To validate test rigor, we injected
artificial errors into ground-truth implementations (like a mutation testing
process); our tests detected 87.04\% of these faults, confirming their
effectiveness at catching subtle mistakes.

\subsubsection{End-to-End ZK Program Generation}
\label{subsubsec:zk-eval-end-to-end}

The final stage assesses LLMs to generate complete and correct ZK programs to
verify computational statements. Building upon the first two stages, it
evaluates the systemic integration of these skills to construct sound,
multi-primitive constraint systems that faithfully encode high-level
verification relations.

\noindent \textbf{Benchmark Design Rationale.}~To ensure a rigorous and
trustworthy assessment in the absence of a standardized ZK-specific generation
dataset, we leverage canonical algorithmic baselines from the well-established
\texttt{HumanEval} dataset. We posit that ZK programming represents a
fundamental paradigm shift from imperative computation to algebraic verification
rather than a shift in the application domain itself. Since the core logic of
production-grade ZK applications (e.g., private finance, decentralized identity,
and state-transition proofs) relies on universal algorithmic patterns (e.g.,
sorting, membership checks, and conditional branching), adapting these general
algorithmic tasks allows us to isolate the gap on computation-to-verification
translation without the confounding variables of application-specific
requirements (e.g., application-specific API calling).

To reflect the unique semantics of ZK proofs, we systematically transform
the original benchmark through three distinct adaptations, as illustrated in the
right column of \F~\ref{fig:zk-eval-overview}:

\begin{enumerate}[leftmargin=*]
    \item \textbf{Compatibility Filtering.} We exclude tasks involving string
    manipulation and floating-point arithmetic which are not supported by DSLs.
    Variable-length output tasks are also excluded since they're incompatible
    with the static nature of arithmetic circuits. Their removal ensures that
    evaluation focuses on the model's constraint-reasoning ability rather than
    DSL limitations. This yields a curated set of 68 standalone tasks suitable
    for ZK synthesis.
    \item \textbf{Verification Reformulation.} We reformulate the natural
    language description of each computational task (defined as finding \( S =
    P(I) \) for input \( I \)) into a verification problem. The LLM must
    generate a verification function $V(I, S) \in \{0, 1\}$ that returns $1$ if
    and only if the candidate solution $S$ satisfies the intended relation for
    input $I$.
    \item \textbf{Oracle Adaptation.} Correspondingly, the ground-truth oracle
    solutions are rewritten from computational functions ($S = P(I)$) to
    verification forms ($V(I, S)$). To ensure these serve as a reliable ground
    truth, we adopt a declarative implementation style; for instance, a sorting
    task is verified by checking permutation and monotonicity predicates rather
    than re-executing a sorting algorithm. To align with the semantics of ZKPs,
    these oracles are restricted to finite field arithmetic (mod $p$) and avoid
    non-deterministic execution paths. Each oracle was manually authored and
    cross-validated against the original computational benchmarks to ensure its
    reliability as a reference for test-case preparation.
\end{enumerate}

\noindent \textbf{Quality Assurance.}~Following established practices in
benchmark construction~\cite{yuan2023no,
siddiq2023exploring,liu2025projecteval,zhang2024naturalcodebench,wang2022recode},
we implement a rigorous validation process to ensure the Soundness and
Completeness of generated programs. While traditional datasets provide only
``Acceptance Cases'' (Completeness), ZK programs require ``Rejection Cases'' to
ensure the circuit is properly constrained (Soundness). Thus we implement
test-case generators to produce input-solution pairs that a correct verifier
must reject, ensuring soundness validation. We cross-validate task descriptions,
adapted oracles, and test-case generators through manual review and injection of
edge cases (e.g., all-zeros, field-overflows, out-of-domain inputs, etc.).
Collectively, this process yields 100 test cases per task with an average
distribution of 54.41\% acceptance and 45.59\% rejection cases. To quantify
benchmark quality, we injected syntactic mutations into the ground-truth oracles
(mutation-testing); our test suites detected 86.42\% mutations, demonstrating
high sensitivity to subtle errors.

\subsection{Measurement Settings}

Using \textsc{ZK-Eval}, we measure LLMs' competencies on ZK code generation. Our
study is guided by the following three research questions (RQs):

\begin{itemize}[leftmargin=*]
    \item \textbf{RQ1: Language \& Toolchain Knowledge.} Do LLMs understand the
    syntax, semantics, and standard library APIs of Circom and Noir?
    \item \textbf{RQ2: Algebraic Primitive Competence.} Can LLMs reliably
    implement and use algebraic primitives, the building blocks of ZK programs,
    to specify correct constraints?
    \item \textbf{RQ3: End-to-End Generation.} Can LLMs integrate these skills
    to produce complete, correct ZK programs on the intended computational
    relation?
\end{itemize}

Each RQ corresponds directly to one component of \textsc{ZK-Eval}. For RQ1,
models are tested on curated multiple-choice questions covering language
knowledge. For RQ2, they are instructed to implement and use algebraic
primitives which are then automatically validated against acceptance and
rejection test cases. For RQ3, they attempt end-to-end program generation to
realize verification constraints for general computational problems, whose
correctness is judged by soundness and completeness against dual test suites.  

\parh{Model Selection.} We evaluate 4 representative LLMs: OpenAI's reasoning
models \emph{GPT-o4-mini} and \emph{GPT-o3}, along with \emph{DeepSeek-V3} and
\emph{Qwen3}. This mix of reasoning and open-weight non-reasoning ones allows us
to assess both family-specific and general trends. All models are run under
identical prompts and budgets, with 10 samples per task to estimate Pass@1 rates
and reduce variance.

\begin{figure*}[t]
    \centering
    \begin{minipage}{.50\textwidth}
        \centering
        \includegraphics[width=\linewidth]{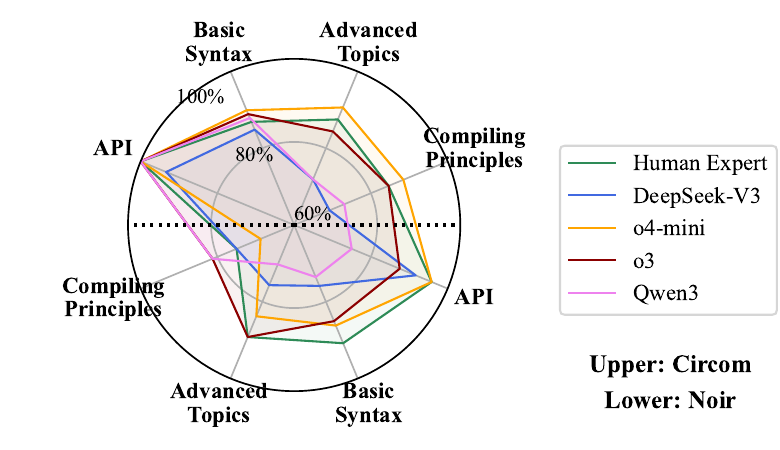}
        % \vspace{-20pt}
        \caption{Accuracy of LLMs and Human Experts on the MCQ Benchmark for ZK Language Knowledge.}
        \label{fig:measurement-results-p1}
    \end{minipage}%
    \hspace{0.02\textwidth}
    \begin{minipage}{.47\textwidth}
        \centering
        \includegraphics[width=\linewidth]{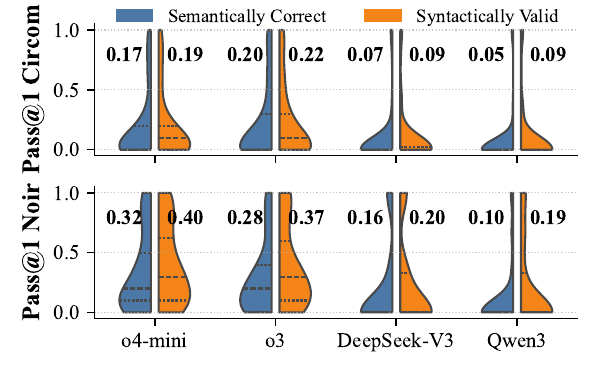}
        % \vspace{-20pt}
        \captionof{figure}{LLM Performance on End-to-end ZK Benchmarks. Annotated with average accuracy.}
        \label{fig:measurement-results-p3}
    \end{minipage}
    \vspace{-10pt}
\end{figure*}
\begin{figure*}[t]
    \centering
    \includegraphics[width=\linewidth]{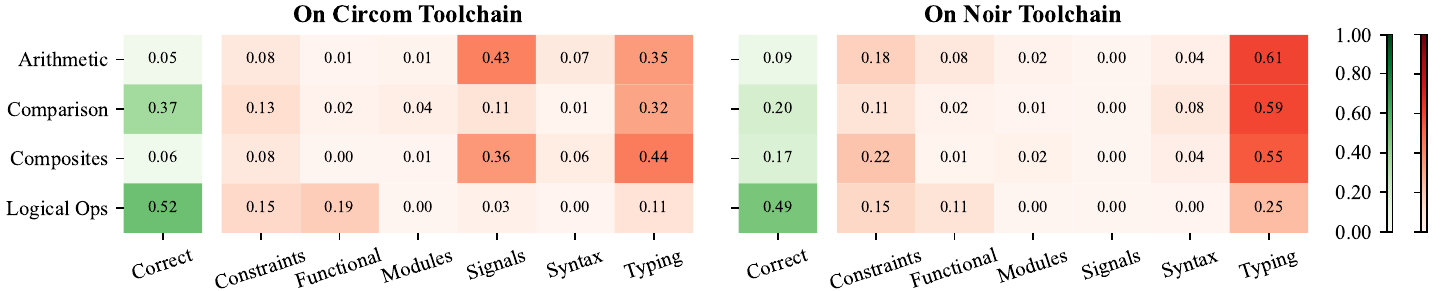}
    \caption{Average Error Distribution of Primitive Implementations Across Languages, Types, and Causes.}
    \label{fig:measurement-results-p2}
    \vspace{-10pt}
\end{figure*}

\subsection{(RQ1) Language \& Toolchain Knowledge}
\label{subsec:rq1}

\F~\ref{fig:measurement-results-p1} summarizes performance. Reasoning models
perform best, with GPT-o4-mini reaching 88.1\% and GPT-o3 87.2\%. Open-weight
non-reasoning models perform noticeably worse, with DeepSeek-V3 at 79.5\% and
Qwen3 at 78.9\%. The human expert for reference scores 88.7\%, similar to the
reasoning models. Overall, LLMs demonstrate strong competence in ZK language
knowledge, achieving accuracy within a few percentage points of the human
expert.

\finding{\ding{172}}{On Circom and Noir, LLMs display strong competence in ZK
language knowledge. Reasoning models surpass open-weight ones and perform
similar to human experts.}

We further examine performance across question categories and languages. LLMs
achieve their highest accuracy on questions about basic syntax and API
references, reflecting familiarity with fundamental constructs and widely used
library templates. In contrast, accuracy drops on advanced topics such as
traits, signal taggings, and generics, which extend expressiveness but are less
commonly encountered. The lowest performance is observed on compiler principles,
which probe understanding of how source programs are transformed into arithmetic
circuits, indicating that they struggle with deeper aspects of arithmetic
circuits and constraint semantics on finite fields. Notably, there is no
significant difference between Circom and Noir, suggesting that model competence
reflects general exposure to ZK languages rather than language-specific
familiarity.

\finding{\ding{173}}{LLMs excel at surface-level knowledge such as syntax and
API references, but exhibit weaknesses in advanced language features and
compiler principles.}

These findings are consistent with the nature of the knowledge base: official
documents, public tutorials, and examples are all widely accessible training
material to LLMs. Thus, it is anticipated that LLMs can achieve strong
performance on such content. This suggests that (i) explicit augmentations on
language grammars or official documentations may be unnecessary (ii) LLMs
exhibit strong potential for ZK programming tasks but require further
evaluations to reveal the bottlenecks.

\subsection{(RQ2) Algebraic Primitive Competence}
\label{subsec:rq2}

Language knowledge enables syntactic fluency, but ZK programming fundamentally
requires the ability to specify algebraic constraints. We next assess whether
LLMs can use or implement algebraic primitives, the atomic elements of
constraint systems. Specifically, model outputs were analyzed with automated
scripts that classified errors based on compiler diagnostics and test outcomes.
From their distinct linguistic and functional effects, we grouped errors into
six categories: (i) \emph{constraint enforcement}, where models failed to encode
relations using equality constructs (\autour{===} in Circom, \autour{assert} in
Noir); (ii) \emph{modules}, involving incorrect instantiation or usage of
library components; (iii) \emph{signal handling}, specific to Circom, where
mistakes occurred in declaring, assigning, or referencing signals; (iv)
\emph{syntax \& parsing}, where the generated code violated grammar rules; (v)
\emph{typing \& resolution}, where data types or variables were used
inconsistently or incorrectly; and (vi) \emph{functional errors}, where code
compiled but failed functional correctness tests.

\F~\ref{fig:measurement-results-p2} shows the results and performance is
consistently weak: even the best-performing category, logical operations,
reached only 52\% accuracy in Circom and 49\% in Noir. Comparison scored lower
(37\% and 20\%), while Arithmetic and Composites dropped to extremely low
accuracy. This contrasts sharply with the strong results on language-level
knowledge: when asked to operationalize this knowledge into specifying
constraints with algebraic primitives, LLMs largely fail to produce testable,
correct implementations. This gap reflects the difficulty of code generation for
low-resource DSLs~\cite{joel2024survey}, exacerbated in ZK programming by the
inherent programming paradigm
shift~\cite{sheybani2025zeroknowledgeproofframeworkssystematic}.

\finding{\ding{174}}{On Circom and Noir languages, LLMs perform consistently
weak on implementing and using algebraic primitives to specify constraints
across all categories.}

Typing \& resolution errors are the most frequent. Noir enforces strict type
matching, disallowing implicit casts between fields and integers, while Circom
distinguishes between variables and signals. Models frequently mix types,
reference undeclared identifiers, or redefine variables, causing compilation
failures. Signal handling errors, unique to Circom, are the next most common,
including declaring signals inside control-flow, reassigning them multiple
times, or using them before initialization. Constraint enforcement errors also
recur: models generate invalid constraint forms such as non-quadratic
constraints (e.g., \autour{x * y * z === 1}) or use off-circuit assertions.
These issues stem from ZK-specific semantics absent in conventional languages.

Performance further varies by primitive type, reflecting library support and
algebraic complexity. Logical operations and comparisons, being boolean
primitives or supported by libraries, are most tractable. Arithmetic and
Composites, such as exponentiation, modulo, floor division, or composites like
\texttt{sum} and \texttt{product}, lack native support and must be implemented
from scratch. These tasks demand reasoning over field arithmetic, careful
handling of inverses and negatives, and explicit soundness constraints. The very
low accuracy here underscores the inability of current models to generate
implementations beyond those directly provided in libraries.

\finding{\ding{175}}{LLMs handle boolean and comparison primitives better than
arithmetic and composite ones, where ZK-specific semantics and limited library
support demand non-trivial reasoning.}

\subsection{(RQ3) End-to-End Generation}
\label{subsec:rq3}

\F~\ref{fig:measurement-results-p3} summarizes the results on the end-to-end
generation benchmark. Compared to language knowledge and primitive-level tasks,
performance drops substantially across all models. For Circom, accuracy is very
low: GPT-o3 reaches only 20\% semantically correct and 22\% syntactically valid
programs, while Qwen3 drops to 5\% and 9\%. Noir results are somewhat higher but
still limited, with GPT-o4-mini at 32\% and 40\% and Qwen3 at 10\% and 19\%.

Several trends emerge from these results. First, Noir tasks consistently yield
higher success rates than Circom, likely due to Noir's simpler type system and
cleaner syntax, which reduce opportunities for errors such as signal misuse or
type mismatches that are pervasive in Circom. Second, reasoning models
substantially outperform open-weight ones: GPT-o4-mini and GPT-o3 are 2--3 times
more successful than DeepSeek-V3 and Qwen3 across both languages, confirming the
model capability difference already observed in earlier tasks. Finally, the gap
between syntactic and semantic correctness underscores that being compilable
alone is insufficient for assessing ZK program quality: many outputs are
compilable but suffer from under- or over-constrained relations.

\finding{\ding{176}}{LLMs perform poorly on end-to-end code generation:
reasoning models beats open-weight ones; Noir is easier than Circom; many
programs compile but remain incorrect.}

This reveals the fundamental challenge of ZK programming: it requires not only
language fluency and primitive competence, but also the ability to architect
multi-primitive constraint systems that faithfully encode complex verification
relations --- a skill that current LLMs have not yet mastered.

%% file: augmenting.tex
\section{Augmenting LLMs for ZK Code Generation}
\label{sec:augmenting}

Based on the findings and gaps revealed from our empirical study in
\S~\ref{sec:capability-study}, we introduce \textsc{ZK-Coder}, an agentic
framework to augment end-to-end ZK code generation. We first summarize the
design principles derived from our findings, then present the system.

\subsection{Design Principles Learned from the Findings}
\label{subsec:findings-to-design}

Findings \ding{172} and \ding{173} show that LLMs possess strong knowledge of ZK
languages' syntax, semantics, and standard library APIs, achieving near
expert-level accuracy. Overall, we interpret that knowledge has been
internalized during contemporary LLM's pretraining stage, eliminating the need
for grammar prompting or documentation access in our augmentation. However,
important challenges remain that motivate our further exploration.

In Findings \ding{174} and \ding{175}, we observe that this knowledge does not
translate into reliable primitive-level generation. Generated code frequently
suffers from typing errors, signal mismanagement, and faulty constraint
enforcement. Finding \ding{176} also reveals these failures in end-to-end
generation. To bridge this gap, \textsc{ZK-Coder} introduces a sketch layer that
decouples abstract constraint modeling from low-level DSL implementation. By
first formalizing relations in this intermediate representation, the framework
can precisely map identified primitives to verified circuit templates and usage
hints via retrieval-augmented generation (RAG)~\cite{gao2023retrieval}. Loosely
speaking, this trades brittle memorization for grounding. Once the logical
``what'' is defined in the sketch, retrieval provides the exact ``how'' for
DSL-specific wiring, collapsing the search space and enabling the reliable
composition of even rare algebraic snippets.

Finally, Findings \ding{174} and \ding{176} highlight end-to-end program
generation as the most challenging setting, where generated programs often fail
due to constraint-misspecification or under/over-constrained logic. These
failures directly undermine the reliability and necessitate mechanisms to
improve validity and faithfulness. To mitigate these failures, \textsc{ZK-Coder}
introduces an interactive refinement loop. The LLM proposes a program, queries
the compiler as an oracle to check constraint specification validity, and, once
passes, self-supplies acceptance/rejection test cases to validate relation
faithfulness. By observing diagnostics and counterexamples, \textsc{ZK-Coder}
repairs its code and iterates the ``\textit{generate-compile-test-repair}'' loop
until these checks pass. This interaction transforms failures into actionable
feedback and guides candidates toward valid, faithful programs.

\subsection{\textsc{ZK-Coder}}
\label{subsec:design-methodology}

\begin{figure*}[t]
    \centering
    \includegraphics[width=0.95\textwidth]{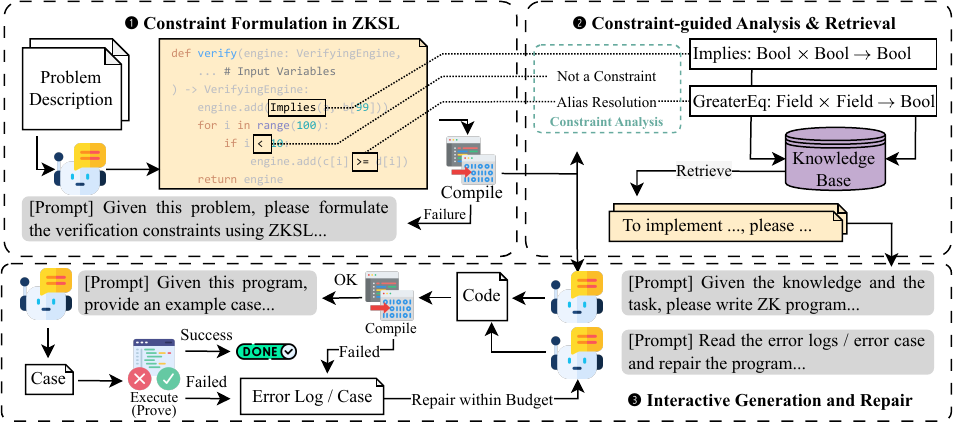}
    \caption{Illustrative Overview on the Design of \textsc{ZK-Coder}.}
    \label{fig:zk-coder-overview}
    \vspace{-6pt}
\end{figure*}

Based on these design principles, we developed \textsc{ZK-Coder}, illustrated in
\F~\ref{fig:zk-coder-overview}. \textsc{ZK-Coder} augments LLMs with a
structured pipeline that bridges natural-language problem descriptions and
executable ZK programs. It consists of three main stages: \ding{192} constraint
formulation, \ding{193} sketch-guided analysis and retrieval, and \ding{194}
interactive generation and repair. 

\subsubsection{Constraint Formulation in Sketch Language.}
\label{subsubsec:design-stage-1}

\input{zksl.tex}

The first stage of \textsc{ZK-Coder} translates natural-language requirements
into a structured specification of verification constraints. Drawing on the
success of planning-based code generation~\cite{wang2024planning,
deng2024assessing, li2023structuredchainofthoughtpromptingcode,
jiang2024selfplanning, wen2025codeplan}, we introduce the ZK Sketch Language
(ZKSL), a Python-embedded Domain Specific Language (eDSL) designed to bridge the
gap between high-level intent and low-level circuit implementation. 

ZKSL is architected to allow LLMs to formalize constraint logic using familiar
Pythonic syntax. This isolates the mathematical ``what'' from the DSL-specific
``how'' by requiring the explicit modeling of constraints over typed inputs to
mirror the verification nature of ZK programs. As formally defined\footnote{We
also provide concrete running examples in the \sm.} in
\F~\ref{fig:zksl-abstract-syntax}, a ZKSL \textsc{Program} is structured around
a \autour{verify} function that accepts these typed inputs and executes a
sequence of constraint-declaration statements via \autour{engine.add($\Phi$)}.
These formulas are composed of relational atoms that correspond to the algebraic
primitives identified in \S~\ref{subsubsec:zk-eval-primitives}. Crucially, all
operations within ZKSL are defined over finite fields (modulo $p$) to faithfully
reflect the mathematical foundation of ZK systems. This enables LLMs to express
complex verification relations using standardized constructs, serving as a
direct logical precursor to the final target code. It allows LLMs to decouple
algebraic reasoning from the implementation-level details (e.g., signal wiring
and constraint encoding) of specific target DSLs.

\textsc{ZK-Coder} prompts the LLM to generate these constraint specification
sketches directly from problem descriptions. To ensure the integrity of the
sketch, the framework performs automated syntactic and type/arity consistency
checks. If a sketch is found to be invalid, the error trace is fed back to the
model for iterative refinement. By serving as a logical representation, ZKSL
decouples the complex task of algebraic constraint reasoning from the low-level
specifics of ZK DSLs (e.g., signal wiring and constraint specification formats),
reducing the overall generation complexity.

\subsubsection{Sketch-guided Analysis and Retrieval.}
\label{subsubsec:design-stage-2}

RAG is a powerful technique for grounding LLM generation in low-resource or
domain-specific contexts~\cite{joel2024survey}. \textsc{ZK-Coder} utilizes the
ZKSL sketch as a structural query to bridge the gap between abstract constraint
specifications and code implementations. Since a ZK program is defined by a set
of verification constraints composed of atomic algebraic primitives, the
framework traverses the sketch to extract each relation declared via
\autour{engine.add($\cdot$)}. To ensure precise retrieval, operators are
canonicalized by normalizing aliases (e.g., \texttt{>=} becomes
\texttt{GreaterEqThan}) and distinguishing variants based on their arity and
field-typing (e.g., differentiating \texttt{XOr} for Booleans vs. Fields). This
stage isolates the constituent units within each constraint to determine the
precise implementation hints required by the target DSL.

To support this stage, we curate a knowledge base of verified implementations
and idiomatic usage patterns for the algebraic primitives defined in
\S~\ref{subsubsec:zk-eval-primitives}. Instead of text-similarity ranking,
\textsc{ZK-Coder} then performs exact-match retrieval for each identified
primitive. These patterns are emitted in the original ZKSL specification order
to preserve the program's logical flow during final code generation. This
structured retrieval effectively grounds the generation process, collapsing the
search space and significantly curbing the model's tendency to hallucinate
incorrect constraint semantics or signal-handling logic.

\subsubsection{Interactive Refinement Loop.}
\label{subsubsec:design-stage-3}

While the sketch provides a logical blueprint, our findings in
\S~\ref{subsec:rq2} and \S~\ref{subsec:rq3} reveal a persistent gap: many
generated programs successfully satisfy DSL grammar rules but fail to faithfully
encode the intended verification relation. Unlike mainstream software where
errors often manifest as runtime crashes, ZK programs frequently yield
under-constrained circuits or over-constrained systems, which silently accept
invalid witnesses or sometimes reject valid ones. To bridge this gap, we
implement an Interactive Refinement Loop as follows.

\begin{enumerate}[leftmargin=*]
    \item \textbf{Constraint Construction Refinement.} The candidate program is
    first submitted to the target compiler (e.g., Circom or Noir) to address
    DSL-specific failures identified in our study, such as signal mismanagement,
    illegal signal declarations in control flow, or constraint-encoding
    violations. We extract these diagnostic traces to iteratively refine the
    constraint wirings until they're valid or reach the retry iteration
    threshold $N_1$ ($N_1$ is set as 8 in our experiments).
    \item \textbf{Relation Faithfulness Refinement.} Once compilable, the
    framework validates constraint faithfulness. The agent generates
    representative ``Acceptance'' and ``Rejection'' test cases to verify that
    the circuit correctly enforces the intended relation, while evaluation test
    cases remain hidden to prevent data-leakage. If any test case fails, the
    counterexample is fed back to re-examine the algebraic constraints. This
    forms another refinement loop, and we set the retry threshold $N_2$ as 3 in
    our experiments.
\end{enumerate}

The above processes repeat until the program passes all functional checks or the
budgets are exhausted, practically converting failures into guided refinements
without manual intervention. 

%% file: zksl.tex
\begin{figure}[t]
\centering
\small
\hrule
\[
    \begin{array}{rrcl}
        \textsc{Program} & P &::=& \texttt{def}\ \texttt{verify}(\texttt{engine},\ x_1\!:\!\tau_1,\ \ldots,\ x_n\!:\!\tau_n):\ S^{*} \\[2pt]
        \textsc{Type} & \tau &::=& \texttt{Field} \mid \texttt{Bool} \mid \texttt{Vec[}\tau\texttt{]} \\[2pt]
        \textsc{Stmt} & S &::=& x = E \mid \texttt{engine.add}( \Phi ) \mid \texttt{for}\ x\ \texttt{in}\ R:\ S^{*} \mid \texttt{if}\ \Phi:\ S^{*}[\texttt{else}:\ S^{*}] \mid \texttt{return engine}\ \\[2pt]
        \textsc{Range} & R &::=& \texttt{range}(E) \mid \texttt{range}(E,E) \\[2pt]
        \textsc{Formula} & \Phi &::=& A_0 \mid \texttt{not}\ \Phi \mid \Phi\ \texttt{and}\ \Phi \mid \Phi\ \texttt{or}\ \Phi \mid \texttt{all}(\Phi\ \texttt{for}\ x\ \texttt{in}\ R) \mid \texttt{any}(\Phi\ \texttt{for}\ x\ \texttt{in}\ R) \\[2pt]
        \textsc{Atom} & A_0 &::=& E\ \mathrel{\bowtie}\ E \mid G(E_1,\ldots,E_n) \\[2pt]
        \textsc{Expr} & E &::=& n \in \mathbb{Z} \mid \texttt{True} \mid \texttt{False} \mid x \mid (E) \mid E\ +\ E \mid E\ -\ E \mid E\ *\ E \mid E\ /\ E \mid E\% E \mid E // E \mid \ldots \\[2pt]
        \textsc{RelOp} & \bowtie &::=& \texttt{==} \mid \texttt{!=} \mid \texttt{<} \mid \texttt{<=} \mid \texttt{>} \mid \texttt{>=} \mid \texttt{and} \mid \texttt{or} \mid \texttt{xor} \\[2pt]
        \textsc{Primitives} & G &::=& \texttt{distinct}(E) \mid \texttt{select}(E, E, E) \mid \texttt{abs}(E) \mid \texttt{sum}(E\ \texttt{for}\ x\ \texttt{in}\ R) \mid \ldots
    \end{array}
\]
\hrule
\caption{The Abstract Grammar of the ZK Sketch Language.}
\label{fig:zksl-abstract-syntax}
\vspace{-10pt}
\end{figure}

%% file: evaluation.tex
\section{Evaluation}
\label{sec:evaluation}

We evaluate \textsc{ZK-Coder} on the task of generating zero-knowledge programs
from natural language specifications. Our evaluation is structured around the
following research questions, with a focus on end-to-end generation. To clarify,
we do not re-evaluate language knowledge (RQ1) or primitive-level coding (RQ2):
the former is established in \S~\ref{subsec:rq1}, and the latter is handled by
our sketching and retrieval mechanism in \S~\ref{sec:augmenting}. Nevertheless,
we include an ablation study (RQ5) to quantify the contribution of each
component.

\begin{itemize}[leftmargin=*]
    \item \textbf{RQ4 (End-to-End Performance):} How does \textsc{ZK-Coder}
    perform compared to baseline methods in generating correct ZK programs?
    \item \textbf{RQ5 (Ablation Study):} What is the effect of each design
    component, including the ZK sketch language, constraint-guided retrieval, and
    the interactive repair loop?
    \item \textbf{RQ6 (Robust Generalization):} How well does \textsc{ZK-Coder}
    generalize to novel, complex computational tasks and real-world
    production-grade ZKP implementations?
    \item \textbf{RQ7 (Failure Analysis):} When \textsc{ZK-Coder} fails, what
    are common failure patterns, and how might they inform future research?
\end{itemize}

\subsection{Experimental Setup}
\label{subsec:experimental-setup}

Our experiments are organized according to the research questions outlined
above.
For \textbf{RQ4 (End-to-End Performance)}, we evaluate on the adapted HumanEval
benchmark from \textsc{ZK-Eval}, using the same four models from our measurement
study. This comparison establishes how well \textsc{ZK-Coder} outperforms
baseline methods in generating correct and provable ZK programs.

For \textbf{RQ5 (Ablation Study)}, we conduct an ablation study on all four
models, selectively disabling each design component of \textsc{ZK-Coder},
including the sketch language, constraint-guided retrieval, and the interactive
repair loop. This aims to quantify their individual impact on overall
performance.  

For \textbf{RQ6 (Robust Generalization)}, we construct a new dataset following
the adaptation methodology of \textsc{ZK-Eval} using LiveCodeBench. Unlike the
simple and potentially contaminated HumanEval, this benchmark contains harder,
contamination-free, and more recent tasks. We evaluate whether \textsc{ZK-Coder}
continues to maintain high performance under more challenging scenarios.

Finally, for \textbf{RQ7 (Failure Analysis)}, we analyze cases where
\textsc{ZK-Coder} fails, categorizing the dominant error patterns and their
underlying causes. This analysis highlights limitations of the current pipeline
and suggests avenues for future research.

All experiments are conducted under a fixed repair budget (see
\S~\ref{subsubsec:design-stage-3}). The evaluation is conducted on a Linux
(Ubuntu 22.04 LTS) server equipped with 256\,GiB of RAM.

\subsection{RQ4: End-to-End Performance}
\label{subsec:end-to-end-performance}

We evaluate \textsc{ZK-Coder}'s end-to-end performance. To the best of our
knowledge, since no prior work directly addresses ZK program generation from
natural language, we establish a simple but meaningful baseline: prompting the
model to generate ZK code directly from task descriptions, with a summary of the
target DSL's grammar and few-shot examples. This setting captures the raw
capability of LLMs as a lower bound. We provide more baselines in
\S~\ref{subsec:ablation-study}.

We meticulously document the performance for the pipeline in three key stages:
(1)~\emph{Sketch correctness} measures whether the initial constraint sketch (in
ZKSL) produced from the natural language description is correct by passing all
hidden test cases; (2)~\emph{Repair pass rate} is conditioned on all correctly
generated sketches, and reports the fraction of these that can be successfully
repaired by the interactive repair loop; (3)~\emph{Program correctness} is
conditioned on all successfully repaired programs, and measures the proportion
that pass all hidden test cases. The final \emph{overall accuracy} aggregates
across the full pipeline, reflecting end-to-end program generation accuracy. In
addition, we report the \emph{average token cost} for each model, which
quantifies the spending.

\input{rq4.tex}

Table~\ref{tab:error-distribution} summarizes the comparative performance of
\textsc{ZK-Coder}. Across all tested models and languages, \textsc{ZK-Coder}
achieves substantial gains over the baseline. On Circom, GPT-o3 leads with
87.85\% accuracy, while open-weight models (DeepSeek V3, Qwen 3) improve from
below 10\% baseline to 49.60\% and 42.43\%, respectively. On Noir, success rates
are higher overall, likely due to its Rust-like syntax being better aligned with
pretraining data. GPT-o3 still leads with 97.79\% accuracy, and GPT-o4-mini
achieves 88.87\%. Open-weight models also show strong gains: Qwen 3 improves
from 11.27\% to 54.74\%, and DeepSeek V3 from 16.18\% to 68.66\%.

The per-stage breakdown highlights the robustness of the pipeline. OpenAI models
show exceptional ``zero-shot'' capabilities in sketching constraints, while
open-weight models exhibit lower correctness rates, suggesting that the leap
from natural language to abstract constraint logic remains a challenge for these
models. The repair loop maintains strong success rates, exceeding 90\% in most
settings, though Qwen 3 on Circom achieves 79.07\%, indicating some difficulty
in recovering from initial errors. Final program correctness is high across most
configurations, though DeepSeek V3 and Qwen 3 show comparatively lower rates on
Circom, suggesting challenges in implementing correct constraint logic. Overall,
the integration of sketching, repair, and semantic validation yields
consistently executable and provable programs across the majority of settings.

We also note that \textsc{ZK-Coder} incurs a higher token cost compared to the
baseline, which is expected since our agentic pipeline includes sketching, RAG
and interactive repair. However, the cost remains modest in practice, with each
task requiring fewer than 5,000 tokens on average, corresponding to less than
0.1 USD per task, making the gains in correctness highly cost-effective.

\answer{4}{\textsc{ZK-Coder} delivers a significant improvement in end-to-end
performance over the baseline method. These results confirm that our pipeline
substantially enhances the performance on LLM-generating ZK programs.}

\subsection{RQ5: Ablation Study}
\label{subsec:ablation-study}

\begin{figure*}[ht]
    \centering
    \includegraphics[width=\textwidth]{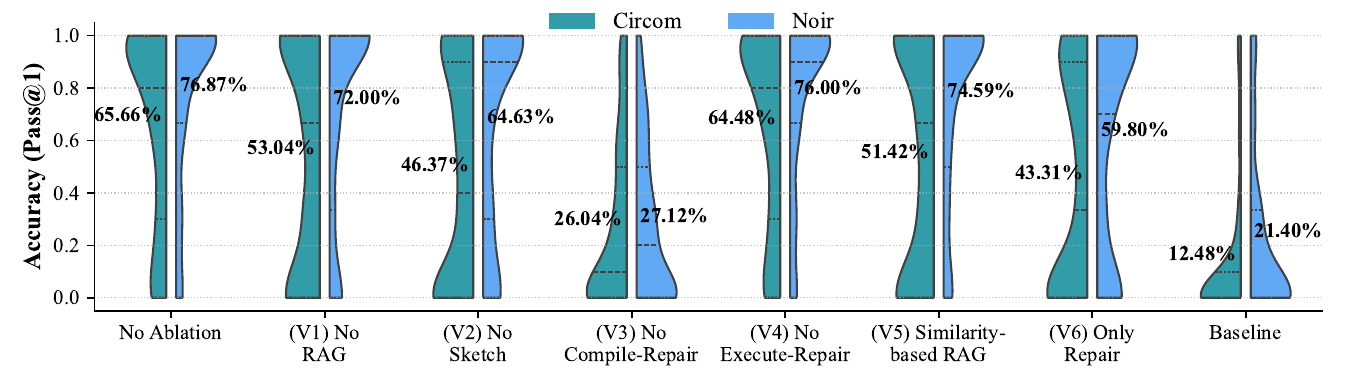}
    \caption{Accuracy Comparison of \textsc{ZK-Coder} Ablation Variants (Averaged Across Models).}
    \label{fig:ablation-study}
\end{figure*}

To assess the role of each major component in \textsc{ZK-Coder}, we conduct an
ablation study where individual modules are disabled or replaced. The following
variants are considered:

\begin{itemize}[leftmargin=*]
    \item \textbf{No RAG (V1):} Disables sketch-guided RAG for algebraic
    primitives, forcing the model to rely solely on its internal knowledge.
    \item \textbf{No Sketch (V2):} Removes the intermediate ZK constraint
    sketching and directly prompts the model to output full programs with RAG
    and the interactive repair loop.
    \item \textbf{No Compile-Repair (V3):} Eliminates the interactive repair
    loop for compiler-identified errors.
    \item \textbf{No Execute-Repair (V4):} Retains compiler-guided repair but
    disables execution and repair.
    \item \textbf{Similarity-based RAG (V5):} Replaces our sketch-guided
    retrieval with a naive text-embedding similarity retrieval on the task
    description.
    \item \textbf{Only Repair (V6):} Omits both retrieval and sketching, relying
    exclusively on test-and-repair.
\end{itemize}

Figure~\ref{fig:ablation-study} presents the Pass@1 accuracy for each ablation
variant across Circom and Noir. The results show that every component plays a
critical role in the overall success of \textsc{ZK-Coder}. Starting from a
full-system average of 65.66\% on Circom and 76.87\% on Noir, we observe that
removing RAG on primitives (V1) causes a significant drop to 53.04\% and 72.00\%
respectively. Disabling the constraint sketching stage (V2) causes further
degradation (46.37\% on Circom, 64.63\% on Noir), highlighting the value of
sketching as a scaffold for task understanding and constraint modeling. The most
severe drop arises from removing compile-repair (V3), with accuracy falling to
26.04\% on Circom and 27.12\% on Noir, demonstrating that interactive repair is
indispensable. Disabling execute-repair (V4) also leads to measurable reductions
(64.48\% on Circom, 76.00\% on Noir), confirming that semantic validation
contributes to robustness. Substituting primitive-in-sketch-guided retrieval
with a text-similarity-based strategy (V5) yields weaker performance (51.42\% on
Circom, 74.59\% on Noir), suggesting that sketch-guided retrieval is more
effective particularly for Circom. Finally, ``Only Repair'' (V6) achieves only
43.31\% on Circom and 59.80\% on Noir, showing that iterative refinement alone
cannot overcome the ``computation-to-verification'' gap. Compared with V6, the
full system achieves an up to 22.35\% performance leap with modest token cost
(20.46\%-29.39\%), confirming that constraint sketching and template retrieval
outperform blind repair cycles.

\answer{5}{The ablation study shows that the strong performance of
\textsc{ZK-Coder} relies on the synergy of sketching, sketch-guided RAG, and
interactive repair. Removing or weakening any of these components causes
noticeable accuracy drops.}

\subsection{RQ6: Robust Generalization}
\label{subsec:rq6-generality}

Our adapted HumanEval serves as a controlled benchmark to evaluate the core
computation-to-verification shift. To evaluate the robustness of
\textsc{ZK-Coder} across varying levels of difficulty and practical contexts, we
extend our analysis to two additional datasets: (i)
LiveCodeBench~\cite{jain2024livecodebench}, which focuses on high-complexity
algorithmic challenges to assess the limits of the framework's reasoning; and
(ii) a set of Production Case Studies curated from 37 open-source ZKP
repositories. To ensure the case studies are representative and avoid selection
bias, we employed a reproducible filtering process targeting high-frequency
circuit patterns such as Merkle verification, state nullification, etc. Details
are provided in the \sm.

\input{rq6.tex}

We establish two baselines: the Simple Baseline (direct generation) and the
Repairing Baseline (adding the interactive repair loop).
Table~\ref{tab:generality-performance} presents the results. On LiveCodeBench,
\textsc{ZK-Coder} achieves a dramatic performance leap, reaching 44.10\% on
Circom and 56.93\% on Noir, significantly outperforming the Repairing Baseline
(14.07\% and 33.95\%). On Production Case Studies, \textsc{ZK-Coder} reaches
near-perfect success rates (90.83\% on Circom and 92.09\% on Noir). While the
Repairing Baseline also performs strongly, this high baseline is expected, as
these studies focus on standardized, commonly used verification patterns in a
controlled context where the high-level logic is well-defined. Nevertheless,
\textsc{ZK-Coder} achieves a noticeable gain by bridging the ``last mile'' of
correctness, reducing subtle field-level constraint errors that persist in the
baselines.

\answer{6}{Results from LiveCodeBench and production case studies confirm that
\textsc{ZK-Coder} generalizes effectively across a spectrum of challenges, from
high-complexity verification logic to standardized production coding patterns.}

\subsection{RQ7: Failure Analysis}
\label{subsec:failure-analysis}

\input{rq7.tex}

Table~\ref{tab:performance-table} reports the distribution of
\textsc{ZK-Coder}'s failure cases. Each column corresponds to a distinct failure
mode. \emph{Repair Budget Exceed} indicates cases where the repair budget was
exhausted without producing any compilable program. \emph{Sketch Incorrect}
denotes that the generated constraint sketch failed hidden test cases,
suggesting an incorrect specification of the verification constraints. \emph{False
Acceptance} corresponds to programs that erroneously accept invalid witnesses,
while \emph{False Rejection} refers to programs that incorrectly reject valid
witnesses. \emph{Mixed False Acc / Rej} captures programs that simultaneously
exhibit both types of errors, reflecting severe constraint mis-specification.

The data reveal that the two dominant sources of failures are \ding{192} and
\ding{193}. In both Circom and Noir, these two together account for over 70\% of
failures. Failure \ding{192} often arises when the model is unable to generate
any compilable program, which is due to gaps in pretraining exposure to ZK-specific
constructs. Failure \ding{193}, on the other hand, reflects errors in correctly
modeling the verification constraints from natural language descriptions. The
less frequent categories (\ding{194}, \ding{195}, \ding{196}) typically stem
from incorrect constraint enforcement when translating a valid sketch into a
concrete circuit. These errors indicate local implementation mistakes rather
than systemic failures of the pipeline.

\answer{7}{In summary, \emph{Repair Budget Exceed} and \emph{Sketched Constraint
Incorrect} constitute the most critical bottlenecks. The former highlights the
need for more ZK-specific training data to improve raw generation capabilities,
while the latter underscores the difficulty of grounding natural language
descriptions in precise verification constraints.}

%% file: rq4.tex
\begin{table}[ht]
    \caption{Comparative Per-stage Performance (Accuracy, Tokens) of \textsc{ZK-Coder} and Baseline.}
    \label{tab:error-distribution}
\resizebox{\textwidth}{!}{
\begin{tabular}{cc|ccc|cc|cc}
\toprule
\multirow{3}{*}{\textbf{Language}} & \multirow{3}{*}{\textbf{Model}} & \multicolumn{3}{c|}{\textbf{\textsc{ZK-Coder} Per-stage Performance}}             & \multicolumn{2}{c|}{\textbf{\textsc{ZK-Coder} Overall}}  & \multicolumn{2}{c}{\textbf{Baseline}} \\
                                   &                                 & (1) \textbf{Sketch}                  & (2) \textbf{Repair} & (3) \textbf{Program} & \textbf{Accuracy}                   & \textbf{Avg.}      & \textbf{Accuracy}  & \textbf{Avg.}    \\
                                   &                                 & \textbf{Correctness}                 & \textbf{Pass Rate}  & \textbf{Correctness} & \textbf{(Pass@1)}                   & \textbf{Token}     & \textbf{(Pass@1)}  & \textbf{Token}   \\ \midrule
\multirow{4}{*}{\textbf{Circom}}   & GPT-o4-mini                     & 94.12\%                              & 91.72\%             & 96.25\%              & 83.38\%                             & 2350.31            & 17.35\%            & 436.98           \\
                                   & GPT-o3                          & 97.30\%                              & 94.41\%             & 95.64\%              & 87.85\%                             & 2155.39            & 20.29\%            & 433.35           \\
                                   & Deepseek V3                     & 72.15\%                              & 92.77\%             & 74.10\%              & 49.60\%                             & 2958.72            & 7.35\%             & 325.28           \\
                                   & Qwen 3                          & 68.12\%                              & 79.07\%             & 78.78\%              & 42.43\%                             & 4979.72            & 4.90\%             & 162.83           \\
\multirow{4}{*}{\textbf{Noir}}     & GPT-o4-mini                     & 94.12\%                              & 97.97\%             & 96.38\%              & 88.87\%                             & 648.95             & 32.21\%            & 152.18           \\
                                   & GPT-o3                          & 97.94\%                              & 99.85\%             & 100.00\%             & 97.79\%                             & 750.63             & 28.38\%            & 191.72           \\
                                   & Deepseek V3                     & 74.88\%                              & 93.36\%             & 94.31\%              & 68.66\%                             & 628.69             & 16.18\%            & 100.91           \\
                                   & Qwen 3                          & 66.84\%                              & 91.88\%             & 89.14\%              & 54.74\%                             & 1260.43            & 11.27\%            & 104.24           \\ \bottomrule
\end{tabular}}
\end{table}

%% file: rq6.tex
\begin{table}[ht]
    \caption{Generality Study on \textsc{ZK-Coder}'s Code Generation Performance (Averaged Across Models).}
    \label{tab:generality-performance}
\centering
\resizebox{\textwidth}{!}{%
\begin{tabular}{c|ccc|ccc}
\toprule
\multirow{2}{*}{\textbf{Language}} & \multicolumn{3}{c|}{\textbf{LiveCodeBench}} & \multicolumn{3}{c}{\textbf{Production Coding Pattern Case Studies}} \\
 & \textbf{\textsc{ZK-Coder}} & \textbf{Repairing Baseline} & \textbf{Simple Baseline} & \textbf{\textsc{ZK-Coder}} & \textbf{Repairing Baseline} & \textbf{Simple Baseline} \\
\midrule
\textbf{Circom} & 44.10\% & 14.07\% & 1.72\% & 90.83\% & 83.33\% & 27.50\% \\
\textbf{Noir}   & 56.93\% & 33.95\% & 9.89\% & 92.09\% & 89.17\% & 31.67\% \\
\bottomrule
\end{tabular}}
\end{table}

%% file: rq7.tex
\begin{table}[ht]
    \caption{Distribution of \textsc{ZK-Coder}'s Code Generation Failures (Averaged Across Models).}
    \label{tab:performance-table}
\small
\resizebox{0.94\textwidth}{!}{
\begin{tabular}{c|cccccc}
\toprule
\multirow{2}{*}{\textbf{Language}} & \textbf{\ding{192} Repair} & \textbf{\ding{193} Sketched}  & \textbf{\ding{194} False Acceptance} & \textbf{\ding{195} False Rejection} & \textbf{\ding{196} Mixed False} \\
                                   & \textbf{Budget Exceed}     & \textbf{Constraint Incorrect} & \textbf{(Under-constrained)}         & \textbf{(Over-constrained)}         & \textbf{Acc / Rej} \\  \midrule
\multirow{1}{*}{\textbf{Circom}}   & 31.67\%                    & 41.56\%                       & 1.08\%                               & 19.78\%                             & 5.92\% \\
\multirow{1}{*}{\textbf{Noir}}     & 12.60\%                    & 73.07\%                       & 1.67\%                               & 10.89\%                             & 1.78\% \\ \bottomrule
\end{tabular}}
\end{table}

%% file: related.tex
\section{Related Work}
\label{sec:related-work}

\paragraph{LLM for DSL Code Generation.}~Emerging research has investigated the
use of LLMs for generating code in DSLs, a task made challenging by limited
training data and specialized syntax~\cite{joel2024survey,
giagnorio2025enhancing}. To mitigate these difficulties, grammar
prompting~\cite{wang2023grammar}, RAG
techniques~\cite{pimparkhede2024doccgendocumentbasedcontrolledcode,
bassamzadeh2024comparative}, iterative refinement~\cite{cycle2024selfrefine},
and knowledge transfer methods~\cite{cassano2024knowledge, mora2024synthetic}
have been proposed. While most of these methods (except knowledge transfer
methods, which target mainstream computational languages) can be adapted for
ZKP code generation, our ablation study (\S~\ref{subsec:ablation-study}) shows
they remain ineffective. \textsc{ZK-Coder} advances this area by designing a
system tailored to the constraint paradigm of ZKPs.

\paragraph{Code Generation Benchmarks.}~Most benchmarks for LLM-based code
generation focus on general-purpose languages~\cite{jain2024livecodebench,
chen2021evaluating}, with limited support for DSLs. A recent
survey~\cite{joel2024survey} reviews existing DSL benchmarks and emphasizes the
lack of standardized evaluation resources, highlighting the need for benchmarks
that capture DSL-specific paradigms, idioms, and practical applications. In the
case of ZKPs, to our knowledge, no systematic benchmark exists for ZKP code
generation that reflects ZK-specific requirements, namely proving a computation,
a static arithmetic-circuit model, and correctness defined by both acceptance
and rejection. We provide a solution for this gap with \textsc{ZK-Eval},
studying code generation bottlenecks and evaluating end-to-end tasks.

\paragraph{Software Engineering for ZKP}~The software engineering community has
recently begun to explore tools and methodologies for ZKP development, with
significant emphasis on security. Static analysis tools such as
Circomspect~\cite{circomspect}, ZKAP~\cite{wen2024practical},
QED$^2$~\cite{pailoor2023automated}, CCC-Check~\cite{kolozyan2025language}, and
halo2-analyzer~\cite{soureshjani2023automated} identify specific vulnerability
patterns, while testing tools \cite{xiao2025mtzk, takahashi2025zkfuzz,
chaliasos2025towards} use fuzzing to expose security issues. These tools are
invaluable for post-hoc analysis and verification, and can be integrated into
our system for further security assurance. Our work complements these efforts by
focusing on the generation phase, leveraging LLMs' capabilities to reduce
engineering burden.

%% file: discussion.tex
\section{Discussion and Limitations}
\label{sec:discussion}

\paragraph{ZKP Program Efficiency.}~Program efficiency for ZKPs is measured by
the size of arithmetic circuits, which directly determines prover cost and
overall performance. To reduce circuit size, a range of compilers have been
proposed to optimize circuit construction~\cite{stronati2024clap, noirlang,
ozdemir2022circ}, while new protocols reduce cost through advanced
arithmetizations with custom gates and lookup
arguments~\cite{chen2023hyperplonk, Gabizon2019PLONKPO}. On the programmer side,
several techniques further improve efficiency, such as adopting ZK-friendly hash
functions~\cite{grassi2021poseidon}, gate packing, and aggregating bits into
single field elements. For LLM-based code generation, however, producing
efficient circuits is more challenging than generating correct ones. As future
work, we plan to survey optimization techniques and guide LLMs on this.

\paragraph{LLM Training and Fine-tuning.}~Recent studies show that LLMs struggle
with DSLs and other low-resource languages due to the lack of high-quality
training data~\cite{joel2024survey}, a challenge that is particularly acute in
the context of ZKPs~\cite{sheybani2025zeroknowledgeproofframeworkssystematic}.
This scarcity also limits the feasibility of fine-tuning approaches. Our work
takes a step toward mitigating this limitation by systematizing the generation
of program examples in ZK DSLs, which can support targeted training and domain
adaptation of LLMs. These examples can serve as a complementary resource for
future research in ZK code generation.

\paragraph{Real-World Representativeness.}~We examine the representativeness and
scope of our evaluation along two complementary dimensions. First, from a
structural standpoint, the programs generated in \S~\ref{subsec:rq6-generality}
exhibit an average cyclomatic complexity of 22.81, which exceeds the average
observed in production-grade open-source ZK repositories (16.8 for Noir and 17.5
for Circom). This indicates that our benchmarks not only mirror but often
surpass the structural depth of real-world ZK logic. Second, while our tasks are
designed to isolate the core challenge of constraint specification, our
evaluation is not ``API-blind.'' We recognize that LLM capabilities in standard
and cryptographic API usage are already well established in prior
work~\cite{qin2023toolllm, zhuo2024bigcodebench, wang2025aicrypto,
zhu2025domaineval}. Consequently, our test suite focuses on verifying that these
APIs are integrated correctly within the circuit logic, covering a considerable
number of real-world in-circuit invocations such as cryptographic primitives and
data-processing utilities. This dual focus ensures that \textsc{ZK-Coder} is
evaluated against the rigorous logical demands of production-level ZK
development.

\paragraph{Threats to Validity.} We identify these potential threats. \ding{192}
Our language-knowledge study uses multiple-choice QA items that were initially
generated with LLM assistance. To mitigate bias and errors, we performed
iterative human validation and cross-checking; while some residual noise may
remain, we expect it to have limited impact on the reported trends. \ding{193}
The \textsc{ZK-Eval} assessment uses hand-crafted test cases that exercise both
accepting and rejecting behaviors of generated programs and primitive usages.
Although tests cannot prove correctness, they include representative corner
cases and explicit checks for soundness and completeness, which substantially
reduce the risk of false positives. The high mutation detection rate reported in
\S~\ref{subsubsec:zk-eval-end-to-end}, achieved through oracle-guided mutation
testing, further supports this claim by demonstrating strong sensitivity to
subtle constraint specification errors. \ding{194} The reliability of our
automated evaluation depends on test-case coverage. We mitigate this through
extensive cross-validation and edge-case injection, achieving a high mutation
score. While this is not a formal proof of logical equivalence, it provides
strong empirical assurance of correctness. \ding{195} The choice of prompting
strategy may influence LLM performance. We employ standard zero-shot and
few-shot prompting techniques without extensive prompt engineering or
fine-tuning. To ensure experimental integrity, we verified that no test cases or
ground-truth solutions were included in the prompts, preventing information
leakage that could artificially inflate success rates. \ding{196} Security is
critical for ZKPs. While integration with auditing
tools~\cite{pailoor2023automated, kolozyan2025language} could further enhance
assurance, code generation from natural language specifications may face
inherent ambiguity and underspecification. Users must exercise caution and
validate generated code before deployment in security-critical application.

%% file: conclusion.tex
\section{Conclusion}
\label{sec:conclusion}

We presented \textsc{ZK-Eval}, the first benchmark for assessing LLM capability
in ZK program generation, revealing that models struggle with composing
algebraic primitives for verification constraints. To address this, we introduce
\textsc{ZK-Coder}, which guides LLMs from constraint sketches to executable code
through sketch-guided retrieval and iterative refinement, significantly
improving code generation for both Circom and Noir.

%% file: illustrative-workflow.tex
\section{Illustrative Workflow Example}
\label{sec:illustrative-example}

\begin{figure*}[ht]
    \centering
    \includegraphics[width=\textwidth]{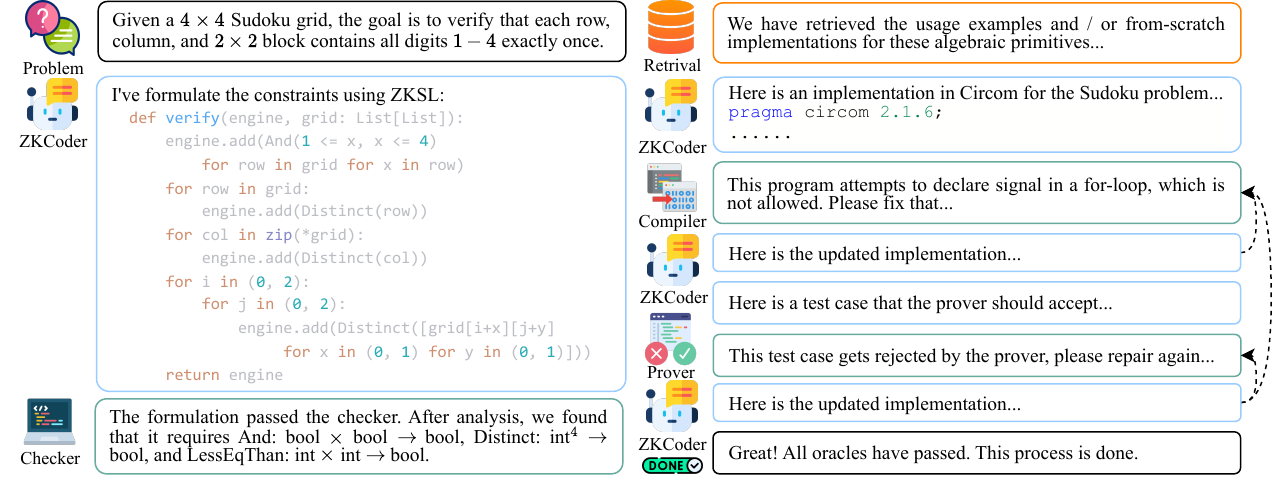}
    \caption{\textsc{ZK-Coder}'s Illustrative Example Workflow on Proving Sudoku Correctness.}
    \label{fig:zk-coder-illustrative-workflow-example}
\end{figure*}

We demonstrate the workflow by using \textsc{ZK-Coder} to generate a ZKP program
for proving the correctness of a Sudoku solution in
\F~\ref{fig:zk-coder-illustrative-workflow-example}, a classical example in ZKP
programming. This example is deliberately chosen for its simplicity and
demonstrative purposes. To clarify, this example is not part of our evaluation
suite. 

The problem is defined as: given a $4\times4$ Sudoku grid, the goal is to verify
that each row, column, and $2\times2$ block contains all digits $1$--$4$ exactly
once.

In this illustrative workflow example, \textsc{ZK-Coder} first reads the problem
description and formulates it in ZKSL, encoding the necessary constraints on the
Sudoku grid. The checker validates the sketch and analyzes it to extract the set
of algebraic primitives required. For each primitive, relevant implementation
hints are retrieved, and an initial Circom/Noir implementation is generated. The
compiler then checks this program for constraint validity, returning diagnostics
that guide interactive refinement. Once the code compiles, the model is prompted
to generate a small test instance (acceptance and rejection). If the resulting
prover behavior deviates from expectations, the model is asked to repair the
program. This repair loop continues until either the program passes the
LLM-supplied tests or the repair budget is exhausted.

\begin{figure*}[htbp]
    \centering
    \includegraphics[width=\textwidth]{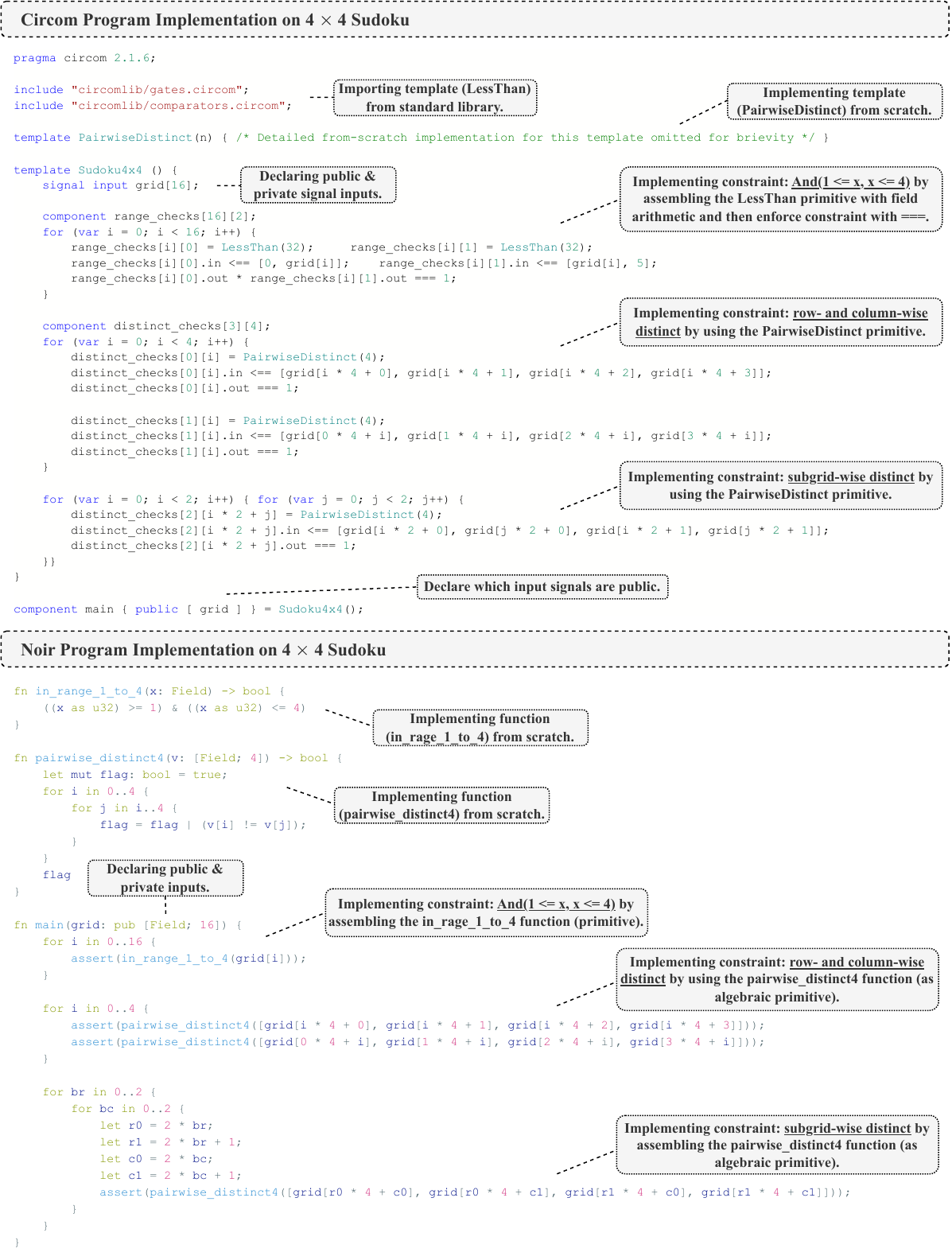}
    \caption{Example Implementation Generated by \textsc{ZK-Coder} (Upper: Circom, Lower: Noir).}
    \label{fig:zk-coder-illustrative-workflow-implementations}
\end{figure*}

We also present two concrete program examples generated by \textsc{ZK-Coder} for
this Sudoku problem in
\F~\ref{fig:zk-coder-illustrative-workflow-implementations}, in Circom and Noir
respectively. Both programs implement the same verification logic. We annotated
the source code with comments to point out key sections, such as input signal
declarations, primitive imports or implementations, and composing algebraic
primitives with field arithmetic operations to implement the constraints
outlined in the ZKSL sketch. For coding tasks on production and
application-specific scenarios, we also provide case studies on frequently used
ZKP coding patterns in production repositories in
\S~\ref{sec:case-studies-public-repo}.

%% file: real-world-coverage.tex
\section{Coverage Study on Algebraic Primitives in Real-World Repositories}
\label{sec:real-world-primitive-coverage}

In this section, we provide the detailed justification and comprehensive listing
of the algebraic primitives used in our study. As detailed in
\S~\ref{subsubsec:zk-eval-primitives}, we constructed a dataset of algebraic
primitives to evaluate the ability of LLMs to specify verification constraints
at the atomic level. These primitives represent the foundational building blocks
of all ZK arithmetic circuits. We present the complete list of the 35 collected
primitives in \T~\ref{tab:algebraic-primitives-list}.

\begin{table*}[htbp]
  \centering
  \caption{List of Collected Algebraic Primitives Used in \textsc{ZK-Eval} and \textsc{ZK-Coder}.}
  \label{tab:algebraic-primitives-list}
  \small
  \begin{tabular}{cccc}
    \toprule
    \textbf{Name} & \textbf{Category} & \textbf{Typing} &  \textbf{Source} \\
    \midrule
    And & Logical & Bool $\times$ Bool $\rightarrow$ Bool & Circomlib \\
    And$^{\ast}$ & Logical & Bool $\times \ldots \times$ Bool $\rightarrow$ Bool & Circomlib \\
    Or & Logical & Bool $\times$ Bool $\rightarrow$ Bool & Circomlib \\
    Or$^{\ast}$ & Logical & Bool $\times \ldots \times$ Bool $\rightarrow$ Bool & Circomlib \\
    XOr & Logical & Bool $\times$ Bool $\rightarrow$ Bool & Circomlib \\
    XOr$^{\ast}$ & Logical & Bool $\times \ldots \times$ Bool $\rightarrow$ Bool & Circomlib \\
    Not & Logical & Bool $\rightarrow$ Bool & Circomlib \\
    Or$^{\diamond}$ & Bitwise & Field $\times$ Field $\rightarrow$ Field & Completeness Aligning \\
    Or$^{\diamond \ast}$ & Bitwise & Field $\times \ldots \times$ Field $\rightarrow$ Field & Completeness Aligning \\
    And$^{\diamond}$ & Bitwise & Field $\times$ Field $\rightarrow$ Field & Completeness Aligning \\
    And$^{\diamond \ast}$ & Bitwise & Field $\times \ldots \times$ Field $\rightarrow$ Field & Completeness Aligning \\
    XOr$^{\diamond}$ & Bitwise & Field $\times$ Field $\rightarrow$ Field & Completeness Aligning \\
    XOr$^{\diamond \ast}$ & Bitwise & Field $\times \ldots \times$ Field $\rightarrow$ Field & Completeness Aligning \\
    Equal & Comparison & Field $\times$ Field $\rightarrow$ Bool & Circomlib, ZK-Kit \\
    NotEqual & Comparison & Field $\times$ Field $\rightarrow$ Bool & Circomlib, ZK-Kit \\
    LessThan & Comparison & Field $\times$ Field $\rightarrow$ Bool & Circomlib, ZK-Kit \\
    LessThanOrEqual & Comparison & Field $\times$ Field $\rightarrow$ Bool & Circomlib, ZK-Kit \\
    GreaterThan & Comparison & Field $\times$ Field $\rightarrow$ Bool & Circomlib, ZK-Kit \\
    GreaterThanOrEqual & Comparison & Field $\times$ Field $\rightarrow$ Bool & Circomlib, ZK-Kit \\
    Add & Arithmetic & Field $\times$ Field $\rightarrow$ Field & ZK-Kit \\
    Subtract & Arithmetic & Field $\times$ Field $\rightarrow$ Field & ZK-Kit \\
    Multiply & Arithmetic & Field $\times$ Field $\rightarrow$ Field & ZK-Kit \\
    Divide & Arithmetic & Field $\times$ Field $\rightarrow$ Field & ZK-Kit \\
    Modulo & Arithmetic & Field $\times$ Field $\rightarrow$ Field & ZK-Kit \\
    Power & Arithmetic & Field $\times$ Field $\rightarrow$ Field & ZK-Kit \\
    FloorDivide & Arithmetic & Field $\times$ Field $\rightarrow$ Field & ZK-Kit \\
    Negate & Arithmetic & Field $\rightarrow$ Field & Circomlib \\
    Absolute & Arithmetic & Field $\rightarrow$ Field & Circomlib \\
    Inverse & Arithmetic & Field $\rightarrow$ Field & Circomlib \\
    Sign & Arithmetic & Field $\rightarrow$ Field & Circomlib \\
    Conditional & Composite & Bool $\times$ Field $\times$ Field $\rightarrow$ Field & Circomlib \\
    Conditional$^{\diamond}$ & Composite & Bool $\times$ Bool $\times$ Bool $\rightarrow$ Bool & Circomlib \\
    Sum$^{\ast}$ & Composite & Field $\times \ldots \times$ Field $\rightarrow$ Field & Circomlib \\
    Product$^{\ast}$ & Composite & Field $\times \ldots \times$ Field $\rightarrow$ Field & Completeness Aligning \\
    Distinct$^{\ast}$ & Composite & Field $\times \ldots \times$ Field $\rightarrow$ Field & Completeness Aligning \\
    \bottomrule
  \end{tabular}
\end{table*}

\begin{table*}[htbp]
  \centering
  \caption{List of Collected Circom (Upper) / Noir (Lower) Repositories on Github.}
  \label{tab:real-world-repo-stats}
  \small
  \resizebox{\textwidth}{!}{
  \begin{tabular}{c|p{9cm}}
    \toprule
    \textbf{Name} & \textbf{Description} \\
    \midrule
    selfxyz/self & A passport authentication system using Circom. \\
    lyronctk/zator & Verified inference of a 512-layer neural network using recursive SNARKs. \\
    FlynnSC/zk-hunt & A prototype for an onchain game, which explores different ZK game mechanics and information asymmetry. \\
    dmpierre/eth-private-market & Circom code using zk-snarks and ethereum for trustlessly selling. \\
    tiktok-privacy-innovation/ & An implementation of trustless attestation verification based on \\
    trustless-attestation-verification-circom & Circom. \\
    nobitex/sigmab & Circom code on Private Proof of Reserves. \\
    erhant/zkbrainfuck & A Brainfuck (programming language) interpreter and virtual machine implemented in Circom. \\
    ava-labs/EncryptedERC & The Encrypted ERC-20 (eERC) standard for secure and confidential token transfers on Avalanche blockchains. \\
    zkFHE/circomlib-fhe & Circom Circuit Library for Fully Homomorphic Encryption. \\
    microsoft/crescent-credentials & Crescent adds privacy to existing credentials with zero-knowledge. \\
    holyaustin/ZeroSecret & A collection of Circom circuits for email verification, hashing, encryption, and more. \\
    inference-labs-inc/subnet-2-circom & source files for the circom implementation of Subnet 2's incentive mechanism. \\
    pluto/aes-circuits & Circom circuits implementing the AES encryption algorithm. \\
    EkuboProtocol/privacy-pools & Circom project implements a privacy pool based on the Privacy Pools Paper. \\
    polymerdao/plonky2-circom & A Plonky2 verifier implemented in Circom. \\
    zkemail/zk-regex & A Circom library for regex matching in zero-knowledge proofs. \\
    0xPolygon/pil-stark & Generates a STARK proof from a State Machine written in PIL Language. \\
    rarimo/passport-zk-circuits & Circom circuits for passport authentication. \\
    nocturne-xyz/protocol & Circom circuits and smart contracts comprising the Nocturne protocol. \\
    \midrule
    AztecProtocol/aztec-nr & The Noir implementation of the Aztec blockchain protocol. \\
    zkemail/zkemail.nr & A privacy-focused email system using zero-knowledge proofs. \\
    colinnielsen/ecrecover-noir & A Noir implementation of the Ethereum ecrecover. \\
    carpalsgrabby/zk-arch-noir-gauge & A Noir example that implements a simple ``architecture style gauge'' for Web3 systems \\
    zkemail/noir-jwt & A Noir library for handling JSON Web Tokens (JWTs) in zero-knowledge proofs. \\
    zac-williamson/mpclib & A library for multi-party computation (MPC) in Noir. \\
    noir-lang/noir\_rsa & A Noir library for RSA encryption and decryption. \\
    distributed-lab/noir-plume & Implementation of the PLUME protocol in Noir. \\
    distributed-lab/op\_rand & A method of emulation of OP\_RAND opcode on Bitcoin. \\
    attested-frontiers/openbanking.nr-circuit & A toolkit for verifying Openbanking payments with the Noir language. \\
    jtriley2p/etherleaks &  Ring Signature-like Protocol for Anonymous Ethereum Signatures. \\
    noir-lang/noir-bignum & A Noir library for big number arithmetic. \\
    hashcloak/noir-mpc-ml & A Noir library for secure multi-party computation (MPC) machine learning. \\
    noir-lang/noir\_string\_search & A Noir library for string searching algorithms. \\
    taurushq-io/private-CMTAT-aztec & Private version of CMTAT security token in Noir. \\
    noir-lang/noir\_json\_parser & A Noir library for parsing JSON data. \\
    noir-lang/noir\_bigcurve & A Noir library that evaluates operations over elliptic curves instantiated with an arbitrary prime field. \\
    defi-wonderland/aztec-standards &  Comprehensive collection of reusable, standardized contracts for the Aztec Network. \\
    \bottomrule
  \end{tabular}}
\end{table*}

Our selection is guided by these principles:
\begin{enumerate}[leftmargin=*]
    \item Theoretical Completeness: These primitives are designed to be a
    functionally complete basis for ZK constraint specification. Since ZK
    programs are ultimately compiled into arithmetic circuits over a finite
    field $\mathbb{F}_p$, any polynomial relation can be decomposed into
    additions and multiplications. By including a broader set of logical,
    relational, and bitwise operators, we ensure that the benchmark can express
    any NP-complete relation.
    \item Expressiveness: While addition and multiplication are sufficient in
    theory, practical ZK development relies on higher-order gates (e.g.,
    multiplexors, range-checks) to manage complexity. Our set includes these to
    mirror this practice.
    \item Excluding Application-Specific APIs: We intentionally exclude
    application-specific APIs (e.g., Merkle tree operations, data-packing, etc.)
    to focus on the core algebraic atomic primitives that underpin all ZK
    programs and are central to constraint specification. Given that said, we
    still provide evaluations on real-world application-specific API calling
    scenarios through case studies on \textsc{ZK-Coder} in
    \S~\ref{sec:case-studies-public-repo}.
\end{enumerate}

\begin{table}[htbp]
\centering
\caption{Coverage and Frequency of Algebraic Primitives Across Repositories.}
\label{tab:real-world-primitive-coverage}
\small
\begin{tabular}{c|cccc}
\toprule
Primitive / Category & Repo Count & Coverage Score & Total Freq. \\ \midrule
Category: Arithmetic & 36 & 0.9730 & 6587 \\
Category: Relational & 35 & 0.9459 & 2428 \\
Category: Logical & 26 & 0.7027 & 547 \\
Category: Composite & 6 & 0.1622 & 9 \\ \midrule
AND & 17 & 0.4595 & 322 \\
OR & 13 & 0.3514 & 32 \\
NOT & 18 & 0.4865 & 103 \\
BITWISE\_AND & 12 & 0.3243 & 34 \\
EQUAL & 34 & 0.9189 & 1771 \\
NOT\_EQUAL & 18 & 0.4865 & 70 \\
LESS\_THAN & 26 & 0.7027 & 218 \\
LESS\_THAN\_OR\_EQUAL & 19 & 0.5135 & 147 \\
GREATER\_THAN & 13 & 0.3514 & 29 \\
GREATER\_THAN\_OR\_EQUAL & 10 & 0.2703 & 40 \\
ADD & 36 & 0.9730 & 1859 \\
SUBTRACT & 32 & 0.8649 & 874 \\
MULTIPLY & 35 & 0.9459 & 2678 \\
DIVIDE & 21 & 0.5676 & 104 \\
MODULO & 20 & 0.5405 & 77 \\
POWER & 12 & 0.3243 & 60 \\
NEGATE & 10 & 0.2703 & 67 \\
CONDITIONAL & 23 & 0.6216 & 153 \\
BITWISE\_XOR & 3 & 0.0811 & 3 \\
XOR & 4 & 0.1081 & 46 \\
BITWISE\_OR & 2 & 0.0541 & 7 \\
INVERSE & 5 & 0.1351 & 18 \\
PRODUCT & 2 & 0.0541 & 2 \\
SUM & 5 & 0.1351 & 7 \\
FLOOR\_DIVIDE & 11 & 0.2973 & 45 \\
\bottomrule
\end{tabular}
\end{table}

To ensure the practical relevance of our primitive set, we conducted a
systematic survey of 37 production-grade ZK repositories (listed in
\T~\ref{tab:real-world-repo-stats}). We collected repositories using GitHub
search for projects implementing Circom or Noir, then filtered out toy,
tutorial, and code-analysis repositories. We further filtered based on
popularity, retaining repositories with at least 30 stars for Circom and 10
stars for Noir, resulting in 19 Circom and 18 Noir repositories. This yielded
408 source files for analysis. 

As shown in \T~\ref{tab:real-world-primitive-coverage}, our primitive set
achieves high coverage. We achieve 97.30\% and 94.59\% coverage for arithmetic
and relational primitives, respectively. These categories form the ``logical
heart'' of nearly all analyzed 408 source files, appearing in 36 out of 37
repositories. Some categories, such as Composite (16.22\%) and specialized
Logical operators (e.g., specific bitwise XORs), show lower coverage. However,
these remain essential for completeness. Overall, this coverage study validates
our primitive selection as practically relevant for evaluating LLM capabilities
in ZK program generation.

%% file: case-studies-public-repo.tex
\section{Case Studies on Frequently Used Production ZK Coding Patterns}
\label{sec:case-studies-public-repo}

Our adapted HumanEval and LiveCodeBench serve as a foundational benchmark to
evaluate the core computation-to-verification shift for ZKP programming. To
evaluate the robustness of \textsc{ZK-Coder} across practical contexts and
domain-specific API using, we extend our analysis with a suite of case studies
derived from 37 production-grade Zero-Knowledge Proof repositories. This section
details the selection criteria, the specific patterns evaluated, and an
elaboration on the performance gains observed.

\subsection{Selection Methodology}

We utilize the same set of 37 production repositories analyzed in
\S~\ref{sec:real-world-primitive-coverage} as the basis for our case studies. To
identify representative ZKP coding patterns, we conduct a systematic review
aimed at isolating logic that is functionally recurrent across real-world
applications. To ensure methodological rigor and mitigate selection bias, we
employ a reproducible processing pipeline:

\begin{itemize}[leftmargin=*]
    \item \textbf{Pattern Extraction:} The authors perform a manual audit of the
    source code to extract coding patterns. We define a ``coding pattern'' as a
    non-trivial logic block that coordinates multiple primitives to enforce a
    high-level architectural invariant. Such patterns typically utilize
    intermediate ``hidden'' signals (witnesses), maintain instance-independence
    (reusability), and compose specific APIs with algebraic constraints to
    describe high-level generalizable verification logic. This collaborative
    audit resulted in an initial set of 269 identified patterns.
    \item \textbf{Clustering and Filtering:} We aggregate these patterns into
    clusters based on their functional roles (e.g., Merkle tree verification,
    state nullification). We exclude patterns that are either secific to a
    single protocol (e.g., Aztec-specific logic in Noir) or functionally trivial
    (e.g., simple API wrappers), or occur only once. This yields a final set of
    10 high-frequency, widely applicable coding patterns, summarized in
    \T~\ref{tab:generalizable-coding-patterns}.
\end{itemize}

\begin{table}[htbp]
\centering
\caption{Frequently Used ZKP Coding Patterns in Production Repositories.}
\label{tab:generalizable-coding-patterns}
\small
\resizebox{\textwidth}{!}{
\begin{tabular}{ccp{10cm}}
    \toprule
    Name & Frequency & Description \\
    \midrule
    Merkle Tree Proof / Verification & 34 & Verify membership of a leaf in a Merkle tree using a path and root hash. \\
    Nullifier Derivation & 16 & Derive deterministic nullifiers from secrets to prevent double-spending. \\
    PubKey / PriKey Consistency & 10 & Verify that a public key is correctly derived from its private key counterpart. \\
    Chunked Data Hashing & 7 & Hash large inputs by decomposing them into smaller fixed-size chunks. \\
    Domain Separated Hashing & 6 & Apply domain separation to hash inputs for different logical contexts. \\
    Commitment Membership Proof & 5 & Verify membership in a set using a commitment-based proof. \\
    Bit-mask Selective Disclosure & 3 & Selectively reveal specific fields using bitmask encoding. \\
    Bitfield Data-packing & 2 & Pack multiple values into a single field using bitfield representation. \\
    ECDH Shared Secret Computation & 2 & Derive shared secrets using Elliptic Curve Diffie-Hellman. \\
    Private Ledger Balance Update & 2 & Update private balances while maintaining ledger consistency. \\
    \bottomrule
\end{tabular}}
\end{table}

\subsection{Evaluation}

\subsubsection{Setup.} To ensure the validity of our results and mitigate the
risk of data leakage (memory recall of production code), we reformulated the
task descriptions to use abstract terminology while maintaining the underlying
algebraic relations. We provide API references alongwith task descriptions for
all baselines to ensure fairness. We follow the experimental setup described in
\S~\ref{sec:evaluation}, using both Circom and Noir as target DSLs with the same
four LLM models. Correctness is evaluated through test cases that are prepared
using the same methodology as in \S~\ref{subsec:zk-eval-design}.

\subsubsection{Results and Analysis.} Results for \textsc{ZK-Coder} on these
case studies are summarized in \T~\ref{tab:generality-performance} (main paper).
We provide detailed analysis on \textsc{ZK-Coder}'s performance for each pattern
below.

\subsection{Case Studies}

We provide case studies on all ten frequently used ZKP coding patterns. For each
case study, we include the problem description, generated ZKSL constraint
sketch, and the final implementation in Circom. Noir is omitted for brevity, but
is available in the replication package.

\input{merkle.tex}
\input{nullifier.tex}
\input{pubkey-privkey.tex}
\input{chunked-hashing.tex}
\input{domain-separated-hashing.tex}
\input{commitment-membership.tex}
\input{bit-mask-selective.tex}
\input{bitfield-packing.tex}
\input{ecdh-secret.tex}
\input{private-ledger-balance.tex}

%% file: merkle.tex
\subsubsection{Merkle Tree Proof / Verification.} Verifying a Merkle path
requires navigating a recursive hash chain while correctly handling left/right
sibling positioning based on a bit-index. It is widely used to prove data
inclusion in a set privately and efficiently. 

Specifically, \textsc{ZK-Coder} first translates the problem into a ZKSL
constraint sketch that captures the core verification logic: compute the Merkle
root via the \autour{ComputeMerkleRoot} API, then enforce conditional equality
between the computed and expected roots using an implication constraint
(\autour{enable}~=~1~$\Rightarrow$~\autour{computed\_root}~=~\autour{root}).
This sketch guides the Circom implementation: the API call is realized through a
\autour{ComputeMerkleRoot} component, while the implication is encoded
algebraically as \autour{(computed\_root[i] - root[i]) * enable === 0},
leveraging field arithmetic to gate the equality constraint. Boolean checks
ensure \autour{enable} and \autour{directions} are binary, completing the
high-level pattern.

\begin{tcolorbox}[breakable, colback=gray!5!white,colframe=gray!50!black,
  colbacktitle=gray!75!black,title=Merkle Tree Proof / Verification] Given a
  Merkle root, a leaf node, and a Merkle proof (path and directions), verify
  whether the leaf belongs to the tree. However, this verification is
  conditional upon a public \autour{enable} flag. 
    
    The process involves:

    \begin{enumerate}[leftmargin=*]
    \item Recomputing the Merkle root from the leaf and the provided siblings using 
       the \autour{ComputeMerkleRoot} API.
    \item If \autour{enable} is 1 (true), the recomputed root must exactly match the 
       provided public root.
    \item If \autour{enable} is 0 (false), the comparison constraint is relaxed, and any 
       recomputed root is accepted.
    \end{enumerate}

    API Syntax:
    
    \autour{ComputeMerkleRoot(leaf, siblings, directions)}: Returns a
    4-element array representing the computed hash root. 
    
    \begin{itemize}[leftmargin=*]
    \item \autour{leaf}: The starting data (array of 4 numbers).
    \item \autour{siblings}: A 2D array [depth][4] of neighbor hashes.
    \item \autour{directions}: A bit array indicating left/right path choices.
    \end{itemize}

    \noindent\rule{\textwidth}{0.4pt}
    
    \begin{verbatim}def verify(
    engine: VerifyingEngine, root, enable, leaf, siblings, directions
) -> VerifyingEngine:
    computed_root = ComputeMerkleRoot(engine, leaf, siblings, directions)
    for i in range(4):
        engine.add(Implies(enable == IntVal(1), computed_root[i] == root[i]))
    return engine
\end{verbatim}
    \noindent\rule{\textwidth}{0.4pt}
    \begin{verbatim}template CircuitProgram() {
    signal input root[4];
    signal input enable;
    signal input leaf[4];
    signal input siblings[12];
    signal input directions[3];
    
    // Boolean constraints for enable and directions
    signal enableBool;
    enableBool <== enable * (enable - 1);
    enableBool === 0;

    signal dirBool[3];
    for (var i = 0; i < 3; i++) {
        dirBool[i] <== directions[i] * (directions[i] - 1);
        dirBool[i] === 0;
    }

    // Compute Merkle root with depth = 3
    component cmr = ComputeMerkleRoot(3);
    for (var j = 0; j < 4; j++) {
        cmr.leaf[j] <== leaf[j];
    }
    for (var k = 0; k < 12; k++) {
        cmr.siblings[k] <== siblings[k];
    }
    for (var d = 0; d < 3; d++) {
        cmr.directions[d] <== directions[d];
    }

    // Conditional equality via gating: (cmr.root[j] - root[j]) * enable == 0
    signal diff[4];
    signal gated[4];
    for (var t = 0; t < 4; t++) {
        diff[t] <== cmr.root[t] - root[t];
        gated[t] <== diff[t] * enable;
        gated[t] === 0;
    }
}
\end{verbatim}
\end{tcolorbox}

%% file: nullifier.tex
\subsubsection{Nullifier Derivation.} 

Nullifier derivation binds a private nullifier and secret to a public commitment
while producing a separate public nullifier hash used for deterministic replay
(double-spend) prevention without revealing the secret nullifier.
\textsc{ZK-Coder} encodes this as a small ZKSL constraint sketch: compute an
intermediate \autour{temp} with \autour{HashTwo(secret, nullifier)}, derive
\autour{commitment} as \autour{HashTwo(temp, amount)}, and produce
\autour{nullifierHash} via \autour{HashOne(nullifier)}. The sketch enforces
equality constraints between the declared outputs and the corresponding hash
results, which map directly to the Circom \autour{HashTwo} and \autour{HashOne}
components in the implementation below. \textsc{ZK-Coder} then retrieves hints
and instructions for this implementation mapping from the sketch to complete the
final code, and the high-level pattern is thus realized.

\begin{tcolorbox}[breakable, colback=gray!5!white,colframe=gray!50!black,
   colbacktitle=gray!75!black,title=Nullifier Derivation] Implement a
  cryptographic commitment and nullifier derivation scheme. The goal is to bind
  a secret value and an amount to a specific nullifier using a multi-stage
  hashing process, while simultaneously generating a public nullifier hash for
  replay prevention.
    
    The process involves:
    
    \begin{enumerate}[leftmargin=*]
    \item Deriving an intermediate \autour{temp} value by hashing a private \autour{secret} 
       and a private \autour{nullifier} using the \autour{HashTwo} API.
    \item Deriving the final public \autour{commitment} by hashing the \autour{temp} value 
       with a public \autour{amount} using the \autour{HashTwo} API.
    \item Deriving a public \autour{nullifierHash} from the private \autour{nullifier} using 
       the \autour{HashOne} API to allow for deterministic double-spend prevention 
       without revealing the nullifier itself.
    \end{enumerate}

    API Syntax:

    \autour{HashOne(value)}: Returns a single field hash of one input.

    \autour{HashTwo(value1, value2)}: Returns a single field hash of two inputs.

    \noindent\rule{\textwidth}{0.4pt}
    
    \begin{verbatim}def verify(
    engine: VerifyingEngine, amount, nullifier,
    secret, commitment, nullifierHash
) -> VerifyingEngine:
    temp = HashTwo(engine, secret, nullifier)
    engine.add(commitment == HashTwo(engine, temp, amount))
    engine.add(nullifierHash == HashOne(engine, nullifier))
    return engine
\end{verbatim}
    \noindent\rule{\textwidth}{0.4pt}
    \begin{verbatim}template CircuitProgram() {
    signal input amount;
    signal input nullifier;
    signal input secret;
    signal input commitment;
    signal input nullifierHash;
    
    // temp = HashTwo(secret, nullifier)
    component h_temp = HashTwo();
    h_temp.a <== secret;
    h_temp.b <== nullifier;
    signal temp;
    temp <== h_temp.out;

    // commitment == HashTwo(temp, amount)
    component h_commit = HashTwo();
    h_commit.a <== temp;
    h_commit.b <== amount;
    commitment === h_commit.out;

    // nullifierHash == HashOne(nullifier)
    component h_nullifier = HashOne();
    h_nullifier.v <== nullifier;
    nullifierHash === h_nullifier.out;
}
\end{verbatim}
\end{tcolorbox}

%% file: pubkey-privkey.tex
\subsubsection{PubKey / PriKey Consistency.} 

Verifying public/private key consistency ensures a private scalar actually
corresponds to a given public point on the Baby Jubjub curve without exposing
the scalar. \textsc{ZK-Coder} encodes this as a ZKSL sketch: decompose the
private key into 253 little-endian bits, iterate those bits and conditionally
add the precomputed base points (where BASE\_POINTS[i] = $2^i * G$) to an
accumulator using the \autour{PointAdd} API, and enforce that the final
accumulator equals the provided public key. The sketch maps directly to Circom
building blocks (e.g., conditional additions are realized arithmetically via
\autour{bit * (sum - acc)} as hinted by \textsc{ZK-Coder}'s retrieved material.)

\begin{tcolorbox}[breakable, colback=gray!5!white,colframe=gray!50!black,
   colbacktitle=gray!75!black,title=PubKey / PriKey Consistency] Verify that a
   private key corresponds to a provided public key on the Baby Jubjub elliptic
   curve (Twisted Edwards form). Instead of using a high-level multiplier, you
   must implement the scalar multiplication logic manually.
    
    The process involves:

    \begin{enumerate}[leftmargin=*]
    \item Decomposing the private key into a 253-bit array (little-endian).
    \item Iterating through the bits. For each bit $i$ that is 1, add the corresponding 
       precomputed base point $P = 2^i * G$ to an accumulator.
    \item The accumulator should start at the curve's neutral element (0, 1).
    \item Use the provided \autour{PointAdd} API to perform the additions. Note that in a ZK 
       circuit, you must handle the conditional addition (adding the point only if 
       the bit is 1) using arithmetic gates rather than branching.
    \end{enumerate}

    Precomputed Base Points:

    A list of 253 points \autour{BASE\_POINTS[i]} is provided, where \autour{BASE\_POINTS[i] = $2^i * G$}.
    
    API Syntax:

    \autour{PointAdd(P1, P2)}: Returns the sum of two points \autour{(x, y)} on the Baby Jubjub curve.

    \noindent\rule{\textwidth}{0.4pt}
    
    \begin{verbatim}def verify(
    engine: VerifyingEngine, senderPublicKey, privKey, BASE_POINTS
) -> VerifyingEngine:
    assert len(senderPublicKey) == 2
    assert len(BASE_POINTS) == 506
    accX = IntVal(0)
    accY = IntVal(1)
    for i in range(253):
        twoPow = IntVal(1 << i)
        bit = (privKey // twoPow) % IntVal(2)
        engine.add(Or(bit == IntVal(0), bit == IntVal(1)))
        Px = BASE_POINTS[2*i]
        Py = BASE_POINTS[2*i+1]
        sx, sy = PointAdd(engine, (accX, accY), (Px, Py))
        accX = accX + bit * (sx - accX)
        accY = accY + bit * (sy - accY)
    engine.add(accX == senderPublicKey[0])
    engine.add(accY == senderPublicKey[1])
    return engine
\end{verbatim}
    \noindent\rule{\textwidth}{0.4pt}
    \begin{verbatim}template CircuitProgram() {
    signal input senderPublicKey[2];
    signal input privKey;
    signal input BASE_POINTS[506];

    // Decompose privKey into 253 bits
    component nb = Num2Bits(253);
    nb.in <== privKey;

    // Accumulators
    signal accX[254];
    signal accY[254];
    accX[0] <== 0;
    accY[0] <== 1;

    // Constraints to ensure each bit is 0 or 1 using OR(bit==0, bit==1)
    component eq0[253];
    component eq1[253];
    component orbit[253];

    // Point addition components and conditional updates
    component pa[253];
    signal deltaX[253];
    signal deltaY[253];
    signal stepX[253];
    signal stepY[253];

    for (var i = 0; i < 253; i++) {
        // Ensure bit is 0 or 1
        eq0[i] = IsEqual();
        eq1[i] = IsEqual();
        orbit[i] = OR();
        eq0[i].in[0] <== nb.out[i];
        eq0[i].in[1] <== 0;
        eq1[i].in[0] <== nb.out[i];
        eq1[i].in[1] <== 1;
        orbit[i].a <== eq0[i].out;
        orbit[i].b <== eq1[i].out;
        orbit[i].out === 1;

        // Point addition: (accX[i], accY[i]) + (Px, Py)
        pa[i] = PointAdd();
        pa[i].x1 <== accX[i];
        pa[i].y1 <== accY[i];
        pa[i].x2 <== BASE_POINTS[2*i];
        pa[i].y2 <== BASE_POINTS[2*i+1];

        // Conditional update: acc = acc + bit * (s - acc)
        deltaX[i] <== pa[i].x3 - accX[i];
        stepX[i] <== nb.out[i] * deltaX[i];
        accX[i+1] <== accX[i] + stepX[i];

        deltaY[i] <== pa[i].y3 - accY[i];
        stepY[i] <== nb.out[i] * deltaY[i];
        accY[i+1] <== accY[i] + stepY[i];
    }

    senderPublicKey[0] === accX[253];
    senderPublicKey[1] === accY[253];
}
\end{verbatim}
\end{tcolorbox}

%% file: chunked-hashing.tex
\subsubsection{Chunked Data Hashing.} 

Chunked data hashing manages large bitstring payloads by partitioning data into
field-compatible segments, a common pattern in identity and large-scale data
attestation. \textsc{ZK-Coder} encodes this logic via a ZKSL sketch that first
mandates boolean integrity checks for the raw \autour{dataBits}, then specifies
the packing of four 190-bit segments into field elements \autour{c0..c3} via the
\autour{BitsToNum} transformation. Finally, the sketch defines a global
commitment constraint where the \autour{dataCommitment} must equal the output of
\autour{HashFive(c0, c1, c2, c3, secretIdentity)}. By retrieving targeted
library implementations for the \autour{Bits2Num} and \autour{HashFive}
primitives, \textsc{ZK-Coder} correctly wires the bit-loop indices and component
inputs, ensuring the circuit accurately realizes the multi-stage hashing
pipeline while maintaining field-capacity safety.

\begin{tcolorbox}[breakable, colback=gray!5!white,colframe=gray!50!black,
  colbacktitle=gray!75!black,title=Chunked Data Hashing] Implement a chunked
  hashing scheme for a large bitstring payload. The goal is to securely commit
  to a long data array (e.g., identity data) by splitting it into chunks that
  fit within the circuit's field capacity.
    
    The process involves:

    \begin{enumerate}[leftmargin=*]
    \item Taking a private bit array \autour{dataBits} of size 760.
    \item Splitting the bits into 4 chunks of 190 bits each.
    \item Converting each 190-bit chunk into a single field element (little-endian).
    \item Passing the 4 derived field elements and a private \autour{secretIdentity} 
       into a hash function to produce a final \autour{dataCommitment}.
    \item Enforcing that the computed hash matches the public \autour{dataCommitment}.
    \end{enumerate}

    API Syntax:

    \autour{BitsToNum(bits)}: Converts an array of bits to a single field element.

    \autour{HashFive(in1, in2, in3, in4, in5)}: Computes a hash of five inputs.

    \noindent\rule{\textwidth}{0.4pt}
    
    \begin{verbatim}def verify(
    engine: VerifyingEngine, dataCommitment, dataBits, secretIdentity
) -> VerifyingEngine:
    assert len(dataBits) == 760
    for b in dataBits:
        engine.add(Or(b == IntVal(0), b == IntVal(1)))
    c0 = BitsToNum(dataBits[0:190])
    c1 = BitsToNum(dataBits[190:380])
    c2 = BitsToNum(dataBits[380:570])
    c3 = BitsToNum(dataBits[570:760])
    h = HashFive(engine, [c0, c1, c2, c3, secretIdentity])
    engine.add(h == dataCommitment)
    return engine
\end{verbatim}
    \noindent\rule{\textwidth}{0.4pt}
    \begin{verbatim}template CircuitProgram() {
    signal input dataCommitment;
    signal input dataBits[760];
    signal input secretIdentity;
    
    // Enforce boolean constraints for each bit: v*(v-1) == 0
    signal bitcheck[760];
    for (var i = 0; i < 760; i++) {
        bitcheck[i] <== dataBits[i] * (dataBits[i] - 1);
        bitcheck[i] === 0;
    }

    // Pack bits into 4 chunks of 190 bits each (little-endian)
    component b2n0 = Bits2Num(190);
    component b2n1 = Bits2Num(190);
    component b2n2 = Bits2Num(190);
    component b2n3 = Bits2Num(190);
    for (var i = 0; i < 190; i++) {
        b2n0.in[i] <== dataBits[i];
        b2n1.in[i] <== dataBits[190 + i];
        b2n2.in[i] <== dataBits[380 + i];
        b2n3.in[i] <== dataBits[570 + i];
    }
    signal c0;
    signal c1;
    signal c2;
    signal c3;
    c0 <== b2n0.out;
    c1 <== b2n1.out;
    c2 <== b2n2.out;
    c3 <== b2n3.out;

    // Compute HashFive over [c0, c1, c2, c3, secretIdentity]
    component hf = HashFive();
    hf.v[0] <== c0;
    hf.v[1] <== c1;
    hf.v[2] <== c2;
    hf.v[3] <== c3;
    hf.v[4] <== secretIdentity;

    // Enforce commitment equality
    hf.out === dataCommitment;
}
\end{verbatim}
\end{tcolorbox}

%% file: domain-separated-hashing.tex
\subsubsection{Domain Separated Hashing.} 

Domain separated hashing is a security-critical pattern that ensures
cryptographic non-malleability by binding a hash result to a specific logical
context. \textsc{ZK-Coder} captures this through a ZKSL sketch that formalizes
the iterative ``absorb-and-permute'' sponge construction: it initializes the
verification state with the \autour{domainSeparator} and recursively computes
intermediate states $s_i$ by summing the previous state with the current
\autour{preimage} element before applying the \autour{MiMC7Permute}
transformation. The sketch explicitly models the data-flow dependencies,
ensuring that each permutation is strictly chained. By retrieving the specific
\autour{MiMC7} round logic and permutation component signatures from its
repository of production templates, \textsc{ZK-Coder} correctly implements the
signal wiring and intermediate summations, ultimately enforcing an equality
constraint between the final permutation output and the public
\autour{outputHash} to guarantee context-specific integrity.

\begin{tcolorbox}[breakable, colback=gray!5!white,colframe=gray!50!black,
  colbacktitle=gray!75!black,title=Domain Separated Hashing] Implement a
  domain-separated hash using a MiMC-7 sponge construction. Domain separation
  prevents the same preimage from producing the same hash across different
  contexts by initializing the hash state with a unique
  \autour{domainSeparator}.
    
    The process involves:

    \begin{enumerate}[leftmargin=*]
    \item Initializing the internal hash state with the public \autour{domainSeparator}.
    \item Absorbing a list of private \autour{preimage} field elements into the state 
       iteratively. 
    \item After each absorption, applying the MiMC-7 permutation rounds.
    \item Constraining the final state to match the public \autour{outputHash}.
    \end{enumerate}

    API Syntax:

    \autour{MiMC7Permute(state)}: Applies the 10-round MiMC-7 permutation to the current field element.

    \noindent\rule{\textwidth}{0.4pt}
    
    \begin{verbatim}def verify(
    engine: VerifyingEngine, domainSeparator, outputHash, preimage
) -> VerifyingEngine:
    s0 = domainSeparator
    s1 = MiMC7Permute(engine, s0 + preimage[0])
    s2 = MiMC7Permute(engine, s1 + preimage[1])
    s3 = MiMC7Permute(engine, s2 + preimage[2])
    engine.add(s3 == outputHash)
    return engine
\end{verbatim}
    \noindent\rule{\textwidth}{0.4pt}
    \begin{verbatim}template CircuitProgram() {
    signal input domainSeparator;
    signal input outputHash;
    signal input preimage[3];

    signal sum0;
    signal sum1;
    signal sum2;

    component p0 = MiMC7Permute();
    component p1 = MiMC7Permute();
    component p2 = MiMC7Permute();

    sum0 <== domainSeparator + preimage[0];
    p0.in <== sum0;

    sum1 <== p0.out + preimage[1];
    p1.in <== sum1;

    sum2 <== p1.out + preimage[2];
    p2.in <== sum2;

    outputHash === p2.out;
}
\end{verbatim}
\end{tcolorbox}

%% file: commitment-membership.tex
\subsubsection{Commitment Membership Proof.} 

The commitment membership proof is a multi-stage verification pattern that
combines cryptographic binding with set inclusion logic. \textsc{ZK-Coder}
models this as a composite ZKSL sketch that simultaneously enforces two distinct
relations: (1) a commitment integrity check, which re-derives the
\autour{positionCommitment} using a \autour{HashThree} of the private
coordinates and nonce; and (2) a disjunctive membership check, which asserts
that the coordinate pair $(x, y)$ matches at least one entry in the public
whitelist \autour{setX} and \autour{setY}. The sketch specifies the use of
parallel equality checks (\autour{IsEqual}) and a logical aggregation tree
(\autour{AND} for coordinate pairs, followed by a multi-input \autour{OR}
chain). \textsc{ZK-Coder} leverages these structural hints to retrieve the
appropriate boolean logic gates and hashing primitives, correctly implementing
the nested loop logic and signal wiring required to prove that the committed
position is authorized without revealing the specific index or the nonce.

\begin{tcolorbox}[breakable, colback=gray!5!white,colframe=gray!50!black,
  colbacktitle=gray!75!black,title=Commitment Membership Proof] Verify that a
  private coordinate pair (x, y) and a private nonce correspond to a public
  \autour{positionCommitment}, and additionally prove that the coordinate pair exists
  within a public whitelist of coordinates.
    
    The process involves:

    \begin{enumerate}[leftmargin=*]
    \item Recomputing a hash of (x, y, nonce) and enforcing that it matches the
       public \autour{positionCommitment}.
    \item Iterating through two public arrays, \autour{setX} and \autour{setY}, to check if 
       there exists an index \autour{i} such that \autour{setX[i] == x} AND \autour{setY[i] == y}.
    \item Enforcing that the membership condition is met (at least one match found).
    \end{enumerate}

    API Syntax:

    \autour{HashThree(val1, val2, val3)}: Computes the hash of three inputs.

    \noindent\rule{\textwidth}{0.4pt}
    
    \begin{verbatim}def verify(
    engine: VerifyingEngine, positionCommitment, setX, setY, x, y, nonce
) -> VerifyingEngine:
    h = HashThree(engine, x, y, nonce)
    engine.add(positionCommitment == h)
    engine.add(
        Or(
            And(setX[0] == x, setY[0] == y),
            And(setX[1] == x, setY[1] == y),
            And(setX[2] == x, setY[2] == y),
            And(setX[3] == x, setY[3] == y),
            And(setX[4] == x, setY[4] == y)
        )
    )
    return engine
\end{verbatim}
    \noindent\rule{\textwidth}{0.4pt}
    \begin{verbatim}template CircuitProgram() {
    signal input positionCommitment;
    signal input setX[5];
    signal input setY[5];
    signal input x;
    signal input y;
    signal input nonce;

    component hash = HashThree();
    hash.v1 <== x;
    hash.v2 <== y;
    hash.v3 <== nonce;

    positionCommitment === hash.out;

    component eqX[5];
    component eqY[5];
    component andXY[5];
    for (var i = 0; i < 5; i++) {
        eqX[i] = IsEqual();
        eqX[i].in[0] <== setX[i];
        eqX[i].in[1] <== x;
        eqY[i] = IsEqual();
        eqY[i].in[0] <== setY[i];
        eqY[i].in[1] <== y;
        andXY[i] = AND();
        andXY[i].a <== eqX[i].out;
        andXY[i].b <== eqY[i].out;
    }

    component or01 = OR();
    or01.a <== andXY[0].out;
    or01.b <== andXY[1].out;

    component or02 = OR();
    or02.a <== andXY[2].out;
    or02.b <== andXY[3].out;

    component or03 = OR();
    or03.a <== or01.out;
    or03.b <== or02.out;

    component or04 = OR();
    or04.a <== or03.out;
    or04.b <== andXY[4].out;
    or04.out === 1;
}
\end{verbatim}
\end{tcolorbox}

%% file: bit-mask-selective.tex
\subsubsection{Bit-mask Selective Disclosure.} 

Bit-mask selective disclosure is a fundamental privacy pattern used to prove the
authenticity of data while revealing only user-authorized fields.
\textsc{ZK-Coder} models this capability through a ZKSL sketch that integrates
three critical verification layers: (1) boolean input validation, which enforces
that the \autour{disclosureSelector} bits are strictly binary to prevent
under-constrained signal manipulation; (2) data integrity binding, which
re-computes the \autour{dataCommitment} from the \autour{privateData} using a
\autour{HashArray} primitive; and (3) conditional masking logic. By sketching
these relations, \textsc{ZK-Coder} ensures that the generated circuit correctly
forces the \autour{revealedData} to zero when the selector is inactive, while
faithfully copying private values when active. The system retrieves the
necessary quadratic constraint templates for the boolean checks, successfully
wiring the multi-signal loop to realize a secure, selective attestation.

\begin{tcolorbox}[breakable, colback=gray!5!white,colframe=gray!50!black,
   colbacktitle=gray!75!black,title=Bit-mask Selective Disclosure] Implement a
   selective disclosure circuit using bitmask selectors. The circuit must prove
   that the \autour{revealedData} array matches a hidden \autour{privateData} array at indices
   where the \autour{disclosureSelector} bit is set to 1, and contains 0 otherwise.
    
    The process involves:

    \begin{enumerate}[leftmargin=*]
    \item Constraining the \autour{disclosureSelector} array to contain only
    binary values (0 or 1).
    \item Validating the integrity of the \autour{privateData} against a public
       \autour{dataCommitment} using hash.
    \item For each index $i$, enforcing the relationship: \\
       \autour{revealedData[i] == privateData[i] * disclosureSelector[i]}.
    \item Proving that if \autour{disclosureSelector[i]} is 0, the corresponding
       \autour{revealedData[i]} is forced to be 0.
    \end{enumerate}
    
    API Syntax:

    \autour{HashArray(array)}: Computes the hash of a field array.

    \noindent\rule{\textwidth}{0.4pt}
    
    \begin{verbatim}def verify(
    engine: VerifyingEngine,
    dataCommitment, disclosureSelector, revealedData, privateData
) -> VerifyingEngine:
    assert len(disclosureSelector) == 4
    assert len(revealedData) == 4
    assert len(privateData) == 4
    for i in range(4):
        engine.add(Or(
            disclosureSelector[i] == IntVal(0),
            disclosureSelector[i] == IntVal(1)
        ))
        engine.add(revealedData[i] == privateData[i] * disclosureSelector[i])
    engine.add(dataCommitment == HashArray(engine, privateData))
    return engine
\end{verbatim}
    \noindent\rule{\textwidth}{0.4pt}
    \begin{verbatim}template CircuitProgram() {
    signal input dataCommitment;
    signal input disclosureSelector[4];
    signal input revealedData[4];
    signal input privateData[4];

    // Enforce disclosureSelector[i] \in {0,1}
    signal sel_minus_one[4];
    signal computedRevealed[4];
    for (var i = 0; i < 4; i++) {
        sel_minus_one[i] <== disclosureSelector[i] - 1;
        disclosureSelector[i] * sel_minus_one[i] === 0;
        // revealedData[i] == privateData[i] * disclosureSelector[i]
        computedRevealed[i] <== privateData[i] * disclosureSelector[i];
        computedRevealed[i] === revealedData[i];
    }

    // Compute and constrain dataCommitment == HashArray(privateData)
    component hasher = HashArray(4);
    for (var i = 0; i < 4; i++) {
        hasher.v[i] <== privateData[i];
    }
    hasher.out === dataCommitment;
}
\end{verbatim}
\end{tcolorbox}

%% file: bitfield-packing.tex
\subsubsection{Bitfield Data-packing.} 

Bitfield data-packing is a fundamental optimization pattern in ZK programming
used to reduce the number of public signals and constraints by aggregating
multiple small values into a single field element. \textsc{ZK-Coder} models this
through a ZKSL sketch that prioritizes algebraic soundness and range integrity.
The sketch defines two distinct sub-proofs: (1) a range-enforcement proof, which
utilizes the \autour{WithinRange} primitive to ensure that \autour{msg\_type}
and \autour{msg\_metadata} do not exceed the 64-bit capacity, preventing
field-overflow vulnerabilities; and (2) an arithmetic reconstruction proof,
which specifies the scaling logic $(msg\_type \cdot 2^{64}) + msg\_metadata$. By
sketching these requirements, \textsc{ZK-Coder} correctly identifies the need
for high-precision comparison components (\autour{GreaterEqThan} and
\autour{LessEqThan}) and constant-time scaling. The system retrieves the exact
templates and hints for range checks, ensuring that the final Circom/Noir code
accurately implements the signals into the target \autour{packedValue} without
silent overflow bugs.

\begin{tcolorbox}[breakable, colback=gray!5!white,colframe=gray!50!black,
   colbacktitle=gray!75!black,title=Bitfield Data-packing] Implement a data
   packing and unpacking scheme using arithmetic scaling. You must pack two
   64-bit unsigned integers (a \autour{type} and \autour{metadata}) into a
   single field element, and then prove that a provided field element correctly
   decomposes into those two specific values.
    
    The process involves:

    \begin{enumerate}[leftmargin=*]
    \item Packing: The \autour{packedValue} is calculated as $(msg\_type \cdot
    2^{64}) + msg\_metadata$.
    \item Range Proof: Enforcing that both \autour{msg\_type} and
       \autour{msg\_metadata} are valid 64-bit integers ($0 \leq x < 2^{64}$).
    \item Unpacking Proof: Enforcing that the public \autour{packedValue}
       exactly matches the result of the arithmetic packing.
    \end{enumerate}

    API Syntax:

      \autour{WithinRange(value, bit\_length)}: Asserts that \autour{value} is
      within the range of $0$ to $2^{bit\_length} - 1$.

    \noindent\rule{\textwidth}{0.4pt}
    
    \begin{verbatim}def verify(
    engine: VerifyingEngine, packedValue, msg_type, msg_metadata
) -> VerifyingEngine:
    engine.add(IntVal(0) <= msg_type)
    engine.add(WithinRange(engine, msg_type, 64))
    engine.add(IntVal(0) <= msg_metadata)
    engine.add(WithinRange(engine, msg_metadata, 64))
    engine.add(packedValue == msg_type * IntVal(2 ** 64) + msg_metadata)
    return engine
\end{verbatim}
    \noindent\rule{\textwidth}{0.4pt}
    \begin{verbatim}template CircuitProgram() {
    signal input packedValue;
    signal input msg_type;
    signal input msg_metadata;

    // Enforce 0 <= msg_type
    component ge1 = GreaterEqThan(252);
    ge1.in[0] <== msg_type;
    ge1.in[1] <== 0;
    ge1.out === 1;

    // Enforce msg_type <= 2^64
    component le1 = LessEqThan(252);
    le1.in[0] <== msg_type;
    le1.in[1] <== 18446744073709551616;
    le1.out === 1;

    // Enforce 0 <= msg_metadata
    component ge2 = GreaterEqThan(252);
    ge2.in[0] <== msg_metadata;
    ge2.in[1] <== 0;
    ge2.out === 1;

    // Enforce msg_metadata <= 2^64
    component le2 = LessEqThan(252);
    le2.in[0] <== msg_metadata;
    le2.in[1] <== 18446744073709551616;
    le2.out === 1;

    // Compute packedValue == msg_type * 2^64 + msg_metadata
    signal high;
    high <== msg_type * 18446744073709551616;
    signal recombined;
    recombined <== high + msg_metadata;
    recombined === packedValue;
}
\end{verbatim}
\end{tcolorbox}

%% file: ecdh-secret.tex
\subsubsection{ECDH Shared Secret Computation.} 

The ECDH (Elliptic Curve Diffie-Hellman) shared secret computation is a
sophisticated cryptographic pattern essential for establishing secure
communication channels within ZK protocols. \textsc{ZK-Coder} handles this task
by drafting a ZKSL sketch that focuses on the algebraic decomposition and scalar
operations required for the Baby Jubjub curve. The sketch defines a two-step
verification process: (1) Bit-decomposition and range validation, which utilizes
the \autour{Num2Bits} primitive to convert the \autour{senderPrivateKey} into a
253-bit array while implicitly enforcing that the scalar fits within the curve's
subgroup order; and (2) Any-Base scalar multiplication, which models the point
multiplication $P_{shared} = d_{sender} \cdot P_{receiver}$. By retrieving
specialized Elliptic Curve templates for \autour{ScalarMulAny} from its curated
repository, \textsc{ZK-Coder} ensures the correct wiring of bit signals to the
scalar multiplication component. This approach guarantees that the resulting
coordinates match the public \autour{sharedKey} without exposing the private
scalar, correctly implementing the complex non-linear constraints of Twisted
Edwards curve arithmetic.

\begin{tcolorbox}[breakable, colback=gray!5!white,colframe=gray!50!black,
   colbacktitle=gray!75!black,title=ECDH Shared Secret Computation] Compute an
   ECDH shared key on the Baby Jubjub curve (Twisted Edwards form). This
   involves performing a scalar multiplication of an arbitrary public point
   (receiverPublicKey) by a private scalar (senderPrivateKey). 
    
    The process involves:
    \begin{enumerate}[leftmargin=*]
    \item Decomposing the private key into a 253-bit array.
    \item Performing Any-Base scalar multiplication: \autour{Result =} \\
    \autour{senderPrivateKey * receiverPublicKey}.
    \item The circuit must verify that the provided \autour{sharedKey} coordinates match the 
       computed point.
    \end{enumerate}

    Note: For this task, you are provided with \autour{PointAdd} and \autour{ScalarMulAny} APIs.

    \noindent\rule{\textwidth}{0.4pt}
    
    \begin{verbatim}def verify(
    engine: VerifyingEngine, receiverPublicKey, sharedKey, senderPrivateKey
) -> VerifyingEngine:
    engine.add(And(
        senderPrivateKey>=IntVal(0),
        senderPrivateKey<IntVal(2**253)
    ))
    bits=[]
    for i in range(253):
        bit=(senderPrivateKey // IntVal(2 ** i)) % IntVal(2)
        engine.add(Or(bit == IntVal(0), bit == IntVal(1)))
        bits.append(bit)
    res_x, res_y = ScalarMulAny(
        engine, (receiverPublicKey[0], receiverPublicKey[1]), bits
    )
    engine.add(res_x == sharedKey[0])
    engine.add(res_y == sharedKey[1])
    return engine
\end{verbatim}
    \noindent\rule{\textwidth}{0.4pt}
    \begin{verbatim}template CircuitProgram() {
    signal input receiverPublicKey[2];
    signal input sharedKey[2];
    signal input senderPrivateKey;
    
    // Decompose senderPrivateKey into 253 bits and enforce range < 2^253
    component bits = Num2Bits(253);
    bits.in <== senderPrivateKey;

    // Scalar multiplication: res = senderPrivateKey * receiverPublicKey
    component scalarMul = ScalarMulAny(253);
    for (var i = 0; i < 253; i++) {
        scalarMul.scalarBits[i] <== bits.out[i];
    }
    scalarMul.point[0] <== receiverPublicKey[0];
    scalarMul.point[1] <== receiverPublicKey[1];

    // Ensure the result equals the sharedKey
    scalarMul.out[0] === sharedKey[0];
    scalarMul.out[1] === sharedKey[1];
}
\end{verbatim}
\end{tcolorbox}

%% file: private-ledger-balance.tex
\subsubsection{Private Ledger Balance Update.} 

The private ledger balance update is a complex state-transition pattern that
encapsulates the core logic of privacy-preserving asset transfers.
\textsc{ZK-Coder} models this through a ZKSL sketch that coordinates three
primary cryptographic objectives: (1) State Consistency, which re-derives the
\autour{oldBalanceCommitment} to prove the user's starting balance without
revealing it; (2) Double-Spend Prevention, which generates a deterministic
\autour{nullifierHash} bound to the specific secret and commitment; and (3)
Arithmetic Integrity, which verifies that the balance subtraction $(new = old -
spend)$ is sound.

The \textsc{ZK-Coder} sketch explicitly mandates a 64-bit range proof using the
\autour{WithinRange} primitive, which effectively enforces the invariant
\autour{oldBalance >= spendAmount}. By retrieving range-check templates,
\textsc{ZK-Coder} correctly wires the intermediate \autour{secretPlusOne}
signals and ensures that the final \autour{newBalanceCommitment} is
algebraically tied to the updated state. This systematic approach allows the
system to realize the full lifecycle of a private transaction while maintaining
strict field-level security.

\begin{tcolorbox}[breakable, colback=gray!5!white,colframe=gray!50!black,
   colbacktitle=gray!75!black,title=Private Ledger Balance Update] Implement a
   private balance subtraction circuit. The circuit proves that a user can spend
   a specific amount from a hidden balance, resulting in a new hidden balance
   and a public nullifier.
    
    The process involves:
    \begin{enumerate}[leftmargin=*]
    \item Recomputing the ``Old Commitment'' by hashing the \autour{oldBalance},
    \autour{secret}, and \autour{nonce}.
    \item Verifying the ``Old Commitment'' matches the public \autour{oldBalanceCommitment}.
    \item Computing a deterministic public \autour{nullifierHash} by hashing the \autour{secret}, 
       the \autour{oldBalanceCommitment}, and a constant padding of \autour{0}. This ensures 
       the nullifier is bound to the specific balance state without revealing 
       the secret.
    \item Calculating \autour{newBalance = oldBalance - spendAmount}.
    \item Enforcing a 64-bit range proof on \autour{newBalance} to ensure the subtraction 
       did not underflow in the prime field (effectively proving \autour{oldBalance >= spendAmount}).
    \item Computing a \autour{newBalanceCommitment} for the remaining funds by
       hashing the \autour{newBalance}, a derived secret (\autour{secret + 1}),
       and the original \autour{nonce}.
    \end{enumerate}

    API Syntax:

      \autour{HashThree(a, b, c)}: Computes a hash of three field elements.

      \autour{WithinRange(value, bit\_length)}: Asserts that \autour{value} is
      within the range of 0 to $2^{bit\_length} - 1$.

    \noindent\rule{\textwidth}{0.4pt}
    
    \begin{verbatim}def verify(
    engine: VerifyingEngine,
    oldBalanceCommitment, newBalanceCommitment, nullifierHash,
    spendAmount, oldBalance, secret, nonce
) -> VerifyingEngine:
    engine.add(HashThree(
        engine, oldBalance, secret, nonce
        ) == oldBalanceCommitment)
    engine.add(HashThree(
        engine, secret, oldBalanceCommitment, IntVal(0)
        ) == nullifierHash)
    newBalance = oldBalance - spendAmount
    engine.add(newBalance >= IntVal(0))
    engine.add(WithinRange(engine, newBalance, 64))
    engine.add(HashThree(
        engine, newBalance, secret + IntVal(1), nonce
        ) == newBalanceCommitment)
    return engine
\end{verbatim}
    \noindent\rule{\textwidth}{0.4pt}
    \begin{verbatim}template CircuitProgram() {
    signal input oldBalanceCommitment;
    signal input newBalanceCommitment;
    signal input nullifierHash;
    signal input spendAmount;
    signal input oldBalance;
    signal input secret;
    signal input nonce;

    // 1) HashThree(oldBalance, secret, nonce) == oldBalanceCommitment
    component hash1 = HashThree();
    hash1.in[0] <== oldBalance;
    hash1.in[1] <== secret;
    hash1.in[2] <== nonce;
    hash1.out === oldBalanceCommitment;

    // 2) HashThree(secret, oldBalanceCommitment, 0) == nullifierHash
    component hash2 = HashThree();
    hash2.in[0] <== secret;
    hash2.in[1] <== oldBalanceCommitment;
    hash2.in[2] <== 0;
    hash2.out === nullifierHash;

    // 3) newBalance = oldBalance - spendAmount
    signal newBalance;
    newBalance <== oldBalance - spendAmount;

    // 4) Ensure newBalance >= 0  <=> spendAmount <= oldBalance
    component lte = LessEqThan(252);
    lte.in[0] <== spendAmount;
    lte.in[1] <== oldBalance;
    lte.out === 1;

    // 5) Ensure newBalance fits in 64 bits
    component rangeCheck = WithinRange(64);
    rangeCheck.in <== newBalance;

    // 6) HashThree(newBalance, secret+1, nonce) == newBalanceCommitment
    signal secretPlusOne;
    secretPlusOne <== secret + 1;
    component hash3 = HashThree();
    hash3.in[0] <== newBalance;
    hash3.in[1] <== secretPlusOne;
    hash3.in[2] <== nonce;
    hash3.out === newBalanceCommitment;
}
\end{verbatim}
\end{tcolorbox}

%% file: filtered-tasks.tex
\section{Task Filtering in Benchmark Curation}
\label{sec:filtered-tasks}

As discussed in the main paper, our end-to-end code generation dataset is
constructed by adapting tasks from HumanEval~\cite{chen2021evaluating} and
LiveCodeBench~\cite{jain2024livecodebench}. These benchmarks were selected for
two primary reasons: first, they are widely recognized as the gold standard for
evaluating code generation, offering a diverse set of tasks that represent
fundamental programming logic; and second, their well-defined computational
problems allow us to isolate and assess the critical
``computation-to-verification'' paradigm shift inherent in ZK programming.

The complete list of these filtered tasks is presented in
\T~\ref{tab:filtered-tasks-human-eval} and
\T~\ref{tab:filtered-tasks-livecodebench}. For brevity, we provide only the task
names and core requirements here; readers interested in the full task
descriptions and implementation details are encouraged to refer to our
accompanying artifact.

Furthermore, for a detailed analysis of our algebraic primitive benchmark,
please see \S~\ref{sec:real-world-primitive-coverage}. For insights and case
studies conducted on production-grade public repositories, refer to
\S~\ref{sec:case-studies-public-repo}.

\begin{table*}[ht]
  \centering
  \caption{Filtered Task IDs from Adapting HumanEval.}
  \label{tab:filtered-tasks-human-eval}
  \small
  \begin{tabular}{ccccccccccccccccc}
    \toprule
    \multicolumn{17}{c}{\textbf{Filtered Task IDs from HumanEval (Total 68 Tasks)}} \\
    \midrule
    3 & 5 & 8 & 9 & 13 & 24 & 31 & 33 & 35 & 36 & 37 & 39 & 40 & 41 & 42 & 43 & 46 \\ 
    49 & 52 & 53 & 55 & 57 & 59 & 60 & 62 & 63 & 65 & 69 & 70 & 72 & 73 & 75 & 76 & 77 \\
    83 & 85 & 88 & 90 & 94 & 97 & 102 & 106 & 108 & 109 & 110 & 114 & 115 & 116 & 121 & 122 & 123 \\
    126 & 127 & 128 & 131 & 135 & 138 & 139 & 142 & 145 & 146 & 147 & 150 & 151 & 152 & 155 & 157 & 159  \\
    \bottomrule
  \end{tabular}
\end{table*}

\begin{table*}[ht]
  \centering
  \caption{Filtered Task IDs from Adapting LiveCodeBench.}
  \label{tab:filtered-tasks-livecodebench}
  \small
  \begin{tabular}{ccccccccccccccccc}
    \toprule
    \multicolumn{17}{c}{\textbf{Filtered Task IDs from LiveCodeBench (Total 34 Tasks)}} \\
    \midrule
    43 & 44 & 62 & 66 & 78 & 94 & 119 & 124 & 129 & 153 & 156 & 172 & 203 & 225 & 232 & 240 & 243 \\ 
    250 & 261 & 265 & 280 & 290 & 297 & 305 & 322 & 323 & 336 & 346 & 353 & 367 & 373 & 375 & 384 & 389 \\
    \bottomrule
  \end{tabular}
\end{table*}

%% file: prompts-available.tex
\section{Prompt Availability Statement}

All prompts used in our experiments are available in the replication package. We
provide exact prompt templates with clearly marked placeholders that are filled
with specific values during \textsc{ZK-Coder}'s execution. 

To ensure methodological fairness, we employed standard zero-shot and few-shot 
prompting techniques without extensive prompt engineering or fine-tuning. 
Additionally, we verified that no test cases or ground-truth solutions were 
included in the prompts, preventing information leakage that could artificially 
inflate success rates.

While we omit raw prompts from this paper to maintain brevity, they are fully 
accessible in the replication package for transparency and reproducibility.